\documentclass[%
 reprint,
superscriptaddress,
 amsmath,amssymb,
 aps,
]{revtex4-2}

\usepackage{graphicx}
\usepackage{dcolumn}
\usepackage{bm}
\usepackage{siunitx}

\usepackage{booktabs}

\begin{document}


\title{Template-Assisted Self Assembly of Fluorescent Nanodiamonds \\ for Scalable Quantum Technologies}

\author{Henry J. Shulevitz}
\affiliation{Department of Electrical and Systems Engineering \\
University of Pennsylvania, Philadelphia PA, 19104, USA}
\author{Tzu-Yung Huang}
\affiliation{Department of Electrical and Systems Engineering \\
University of Pennsylvania, Philadelphia PA, 19104, USA}
\author{Jun Xu}
\affiliation{Department of Electrical and Systems Engineering \\
University of Pennsylvania, Philadelphia PA, 19104, USA}
\author{Steven Neuhaus}
\affiliation{Department of Materials Science and Engineering\\
University of Pennsylvania, Philadelphia PA, 19104, USA}
\author{Raj N. Patel}
\affiliation{Department of Electrical and Systems Engineering \\
University of Pennsylvania, Philadelphia PA, 19104, USA}
\author{Lee C. Bassett}
\email[Corresponding authors. ]{lbassett@seas.upenn.edu \& kagan@seas.upenn.edu }
\affiliation{Department of Electrical and Systems Engineering \\
University of Pennsylvania, Philadelphia PA, 19104, USA}
\author{Cherie R. Kagan}
\email[Corresponding authors. ]{lbassett@seas.upenn.edu \& kagan@seas.upenn.edu }
\affiliation{Department of Electrical and Systems Engineering \\
University of Pennsylvania, Philadelphia PA, 19104, USA}
\affiliation{Department of Materials Science and Engineering\\
University of Pennsylvania, Philadelphia PA, 19104, USA}
\affiliation{Department of Chemistry\\
University of Pennsylvania, Philadelphia PA, 19104, USA}
  
\date{\today}
             

\def \Measurements {\begin{figure*}
\includegraphics[width = \textwidth ]{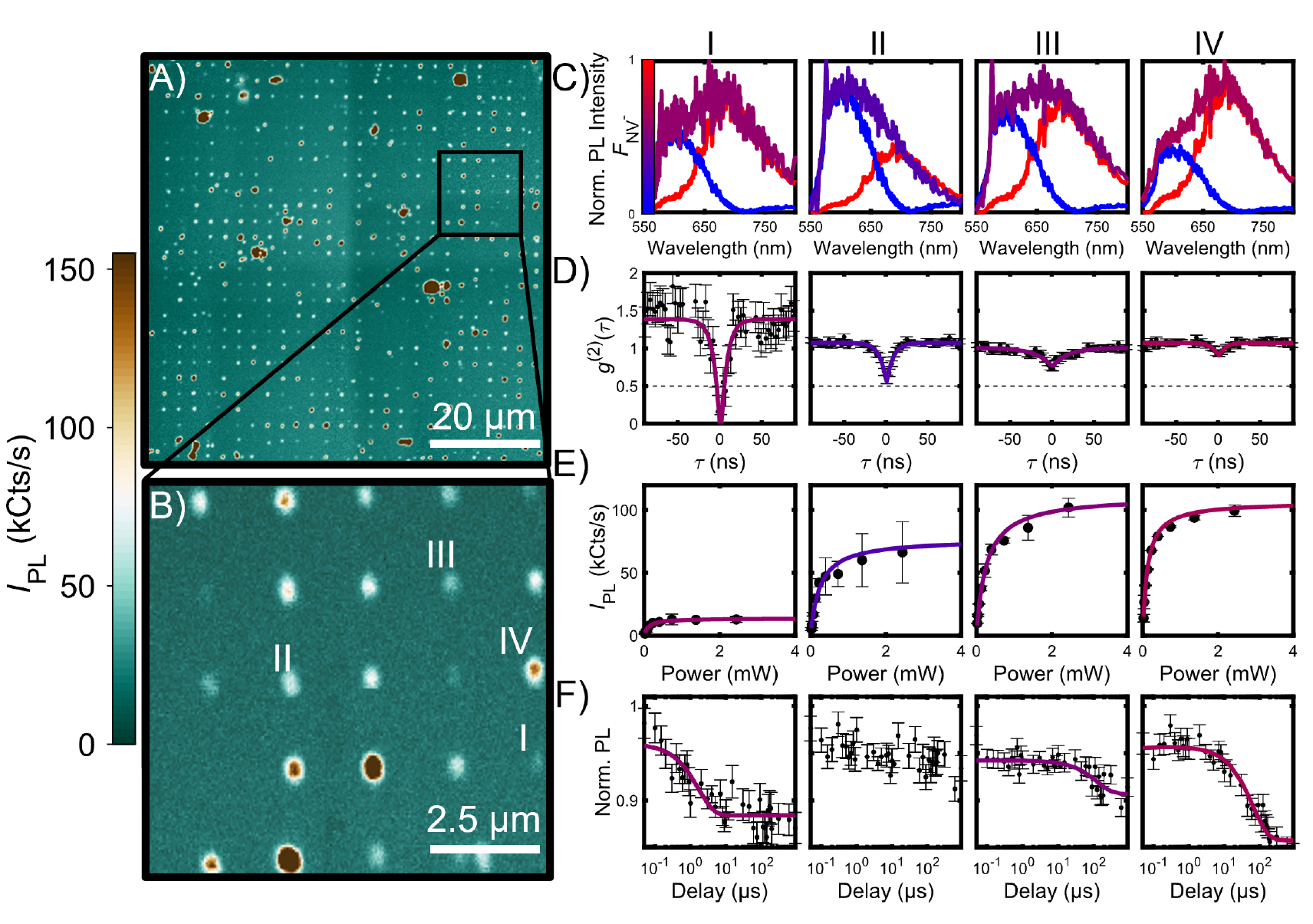}
\caption{\label{fig:Automation Overivews}
 \textbf{Automated characterization of assembled nanodiamonds.} 
 (A) PL scan of the nanodiamond array shown in Figure~\ref{fig:Yields}A. 
 (B) Higher-magnification PL scan of the 5$\times$5 subarray indicated by a black box in (A). Labels I-IV indicate individual nanodiamonds featured in panels C-F. 
 (C) PL spectra of individual nanodiamonds I-IV and their decomposition into NV$^{-}$emission (red) and NV$^{0}$ emission (blue). The overall spectra are color coded to represent the charge ratio, ($F_\mathrm{NV^{-}}$). This color code is maintained for fits in panels D-F. 
 (D) Autocorrelation, (E) PL saturation, and (F) spin relaxation time ($T_1$) measurements of nanodiamonds I-IV (black points), with corresponding fits to the models described in the text (colored curves). Error bars represent experimental uncertainties due to shot noise and slow signal variations. 
}
\end{figure*}}

\def \Histograms {\begin{figure}[t]
\centering
\includegraphics[scale=1]{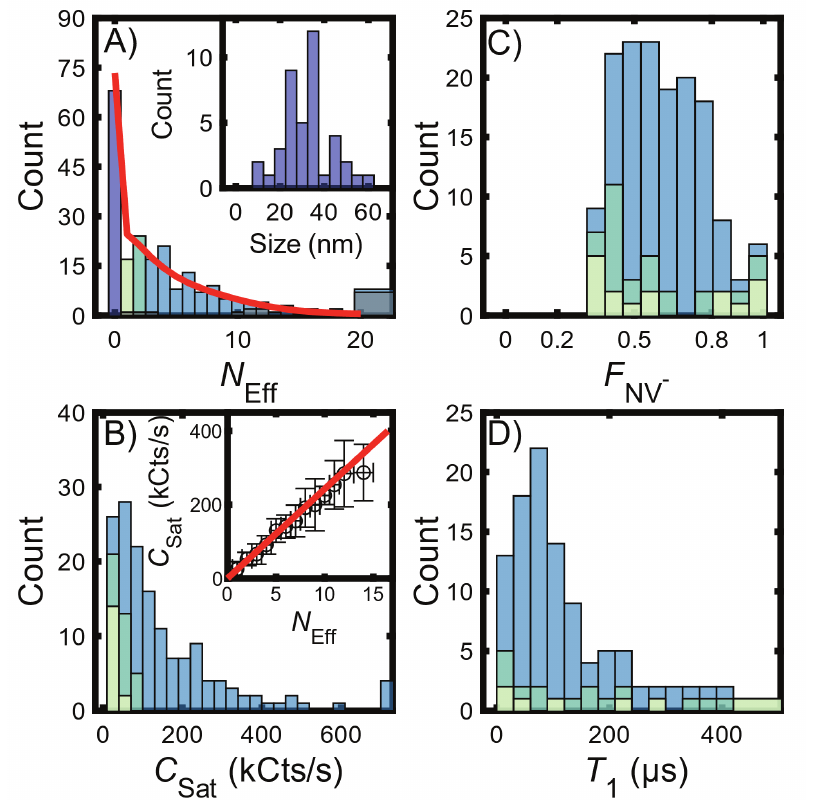}
\centering
\caption{\label{fig:Data Overivews}
\textbf{Optical and quantum properties of nanodiamonds.}
 Throughout the figure, color-coded histograms indicate nanodiamonds with ${N}_\mathrm{Eff} = 0$ (purple), $0< N_\mathrm{Eff}<1.5$ (yellow-green), $1.5< N_\mathrm{Eff}<2.5$ (green), and ${N}_\mathrm{Eff}>2.5$ (blue).
(A) Distribution of ${N}_\mathrm{Eff}$, calculated from autocorrelation measurements (colored bars) or estimated from ${C}_\mathrm{Sat}$ (grey bars). 
The red curve is the result of a fit to the model described in the text.
(Inset) AFM height distribution for the non-fluorescent nanodiamonds only.
Panels (B), (C), and (D) respectively show the measured distributions of ${C}_\mathrm{Sat}$, $F_\mathrm{NV^{-}}$, and $T_1$, respectively, for the ensemble of fluorescent nanodiamonds.  The inset to panel (B) shows ${C}_\mathrm{Sat}$ as a function of ${N}_\mathrm{Eff}$ (data points), along with a linear fit (red line).
}
\end{figure}}

\def \ProjectOverview {\begin{figure}[b]
\centering
\includegraphics[scale=1]{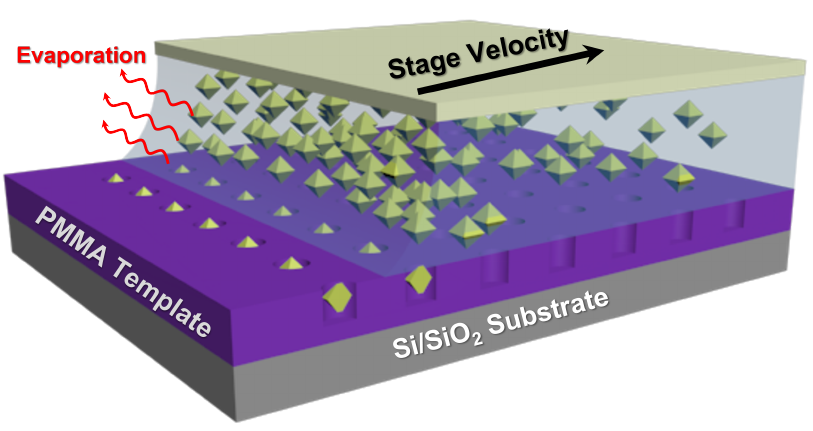}
\centering
\caption{\label{fig:Project Overivew}
\textbf{The TASA method.} Schematic of the capillary-driven, template-assisted self assembly (TASA) of milled, fluorescent nanodiamonds into lithographically defined, PMMA templates on a Si/SiO$_2$ substrate.
} \end{figure}}

\def \Yields{\begin{figure}[t]
\centering
\includegraphics[scale=1]{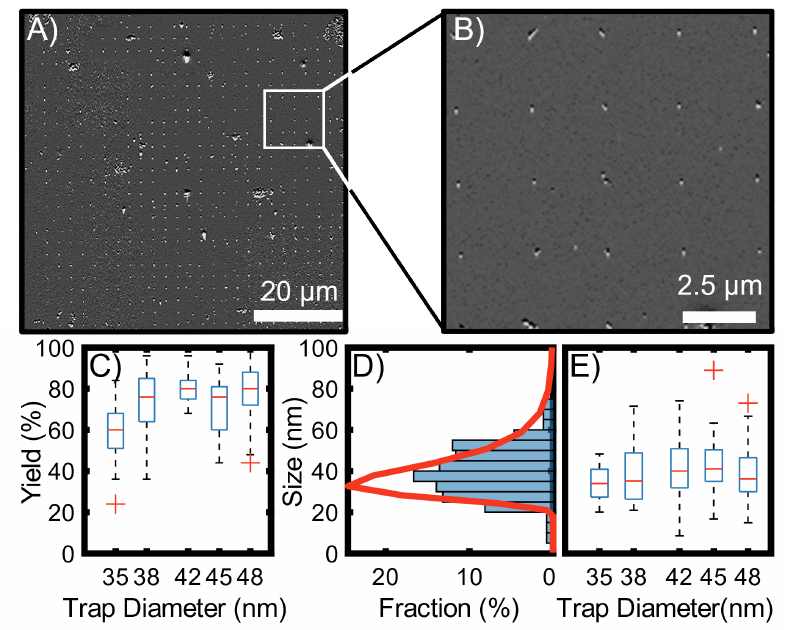}
\centering
\caption{\label{fig:Yields}
\textbf{Assembly yield and size distribution.} (A) Stitched AFM images of an assembly based on a 26$\times$26 array of 42-nm-diameter trap sites.
(B) Higher resolution AFM image of the 5$\times$5 subarray indicated by a white box in (A). The lateral dimensions of the individual nanodiamonds are enlarged due to convolution with the AFM probe tip. 
(C) Assembly yield \textit{vs.} trap diameter. The box plot statistics  are calculated by dividing the array into 5$\times$5 nanodiamond subarrays to account for regional variations. 
(D) Histogram of AFM-measured, assembled nanodiamond heights over all templates (258 individual nanodiamonds) and a DLS measurement (red line) of the parent nanodiamond dispersion.
(E) Nanodiamond size distribution as a function of template trap diameter. 
Outliers (red crosses) in (C) and (E) indicate measurements whose values are $>$1.5 times the interquartile range from the bottom or top of the box.
} \end{figure}}
\begin{abstract}
 Milled  nanodiamonds containing nitrogen-vacancy (NV) centers  provide an excellent platform for sensing applications as they are optically robust, have nanoscale quantum sensitivity, and form colloidal dispersions which enable bottom-up assembly techniques for device integration.
However, variations in their size, shape, and surface chemistry limit the ability to position individual nanodiamonds and statistically study properties that affect their optical and quantum characteristics.  
Here, we present a scalable strategy to form ordered arrays of nanodiamonds using capillary-driven, template-assisted self assembly.  
This method enables the precise spatial arrangement of isolated nanodiamonds with diameters below 50 nm across millimeter-scale areas.
Measurements of over 200 assembled nanodiamonds yield a statistical understanding of their structural, optical, and quantum properties.
The NV centers' spin and charge properties are uncorrelated with nanodiamond size,
but rather are consistent with heterogeneity in their nanoscale environment. 
This flexible assembly method, together with improved  understanding of the material, will enable the integration of nanodiamonds into future quantum photonic and electronic devices.
\end{abstract}

\maketitle{}

\section{\label{sec:level1}Introduction}
The nitrogen-vacancy (NV) center in diamond functions as an optically-addressable spin qubit with room-temperature spin coherence and sensitivity to environmental perturbations such as magnetic and electric fields, strain, temperature, and pressure \cite{doherty2013nitrogen,rondin2014magnetometry,schirhagl2014nitrogen,hopper2018spin}.
Although the best spin coherence properties are achieved in bulk diamond crystals \cite{balasubramanian2009ultralong, sangtawesin2019origins}, nanodiamond particles with diameters $<$ 100 nm still shield the spin qubit from decoherence \cite{tetienne2013spin}, and unlike bulk diamond, they can be dispersed in solvents to form colloids and placed within nanoscale distances of biological or inorganic structures.
As an optically robust, biocompatible, quantum sensitive colloid, nanodiamonds provide a platform for sub-diffraction limited imaging,  \textit{in vivo}  and  \textit{in vitro} sensing, and integration with complex heterojunction devices \cite{ taylor2008high, schirhagl2014nitrogen, krueger2011beyond}. 

Two methods are typically used to prepare nanodiamonds: detonation synthesis and milling bulk crystals. 
Detonation nanodiamonds are useful as dyes and biomarkers; they offer good size uniformity, but the outer layer of sp$^2$ carbon and high background impurity levels associated with their synthesis inhibit the formation and stability of NV centers \cite{schirhagl2014nitrogen, mochalin2012properties}.
On the other hand, milled nanodiamonds inherit the chemical purity of their source crystal, at the cost of uniformity \cite{nagl2015improving}.
Milled nanodiamonds vary widely in size, shape, surface chemistry, and number of NV centers \cite{schirhagl2014nitrogen}.
These chemical and structural variations are reported to give rise to inhomogeneity in the nanodiamonds' optical and quantum properties \cite{tetienne2013spin, ong2017shape, reineck2019not}. 
Statistical studies of individual, isolated nanodiamonds are therefore necessary to correlate their structure and properties and ultimately to optimize nanodiamond quantum devices.

The variation in nanodiamond size typically exceeds 10-50$\%$ depending on the milling parameters \cite{ong2017shape,shenderova2017commercial, reineck2019not}.
This variation, in addition to heterogeneity in shape and surface chemistry hinders the placement or assembly of individual particles and their incorporation into devices. 
In past studies, researchers have used atomic force microscopes (AFMs) to serially position individual nanodiamonds \cite{shafiei2013subwavelength, shi2014modular, sun2016interplay, bogdanov2019hybrid},
or chosen to forgo ordered assemblies and search for randomly dispersed particles \cite{bogdanov2017electron, ong2017shape, hopper2018amplified, reineck2019not}.
Self-assembly techniques have been employed to assemble clusters of several nanodiamonds \cite{ heffernan2017nanodiamond}.
Nanodiamond placement has been achieved through electrostatic interactions between the particles and specific substrates \cite{jiang2015reusable}.
These approaches however place limits on the type of templates that can be used and hinder large statistical studies of individual nanodiamonds.

Here, we employ capillary-driven, template-assisted, self assembly (TASA) to form millimeter-scale area arrays of individual, milled nanodiamonds. 
TASA combines the flexibility and precision of nanolithography with the scalability of colloidal assembly.
It has been used to position individual nanoparticles, as well as to direct the orientation of anisotropic particles, of various materials \cite{greybush2014plasmon, greybush2017plasmon, malaquin2007controlled, mehraeen2015directed, fringes2019deterministic, asbahi2015large, ni2018capillary, preuss2021assembly}.
By using these arrays and an automated optical measurement system, we study the statistical heterogeneity of the nanodiamonds' structural, optical, and spin properties. 
Our studies uncover statistical variations in nanodiamond properties correlated with characteristics of the underlying diamond material, the nanodiamond creation process, and the NV centers' local environment.
This understanding will inform the realization of improved nanodiamond materials, and the versatile TASA method can be utilized in fabricating devices for quantum sensing and quantum photonics.

\section{Results and Discussion}

\subsection{Assembly Process}
We utilize commercially-available, low-fluorescence, milled nanodiamonds from Ad\'amas  Nanotechnologies. 
The nanodiamonds are specified to contain an average of 1-4 NV centers (with approximately 13$\%$ containing a single NV center), have an average size of $\sim$ \SI{40}{\nano\meter}, and have surface carboxylate anions that enable their stable colloidal dispersion \cite{shenderova2017commercial, https://www.adamasnano.com/}.

Figure~\ref{fig:Project Overivew} depicts the TASA process. 
Electron-beam lithography is used to fabricate templates with cylindrical traps, ranging in design diameter from 32-\SI{200}{\nano\meter} and having heights of \SI{62}{\nano\meter}, in poly-(methyl methacrylate) (PMMA) thin films deposited on \SI{250}{\nano\meter} SiO$_2$ on Si substrates.
Larger trap diameters ($>$ \SI{50}{\nano\meter}) result in multiple particle assembly (Supporting Information Figure ~S1) and smaller templates ($<$ \SI{35}{\nano\meter}) show limited particle assembly.
Here, we focus on 35-\SI{48}{\nano\meter} trap diameters with a \SI{2.6}{\micro\meter} pitch to simplify single-nanodiamond spectroscopic and morphological  characterization, and we fabricate 26$\times$26 nanodiamond arrays.

\ProjectOverview{}

To assemble the nanodiamonds, we use a home-built apparatus in which an aqueous nanodiamond dispersion is deposited between a glass slide and the template surface (Figure \ref{fig:Project Overivew}).
A motororized stage translates the nanodiamond dispersion at a velocity of \SI{3.5}{\micro\meter}/s across the template surface.
An accumulation of particles forms at the meniscus and capillary forces drive the nanodiamonds into the trap sites \cite{greybush2014plasmon, yin2001template, mehraeen2015directed, fringes2019deterministic, asbahi2015large}.
The assembly apparatus is enclosed in an environmental chamber to maintain a dew point of 8.6-9.5 $^{\circ}$C. 
A recirculating chiller and heater is used to pump water through a copper block that serves as a sample holder and
maintains a substrate temperature of 24-25  $^{\circ}$C.
Drier chamber conditions yield assemblies with excess nanodiamond deposition, overfilling the trap sites.
A wetter chamber results in limited assembly.
After assembly, the PMMA template is removed by sequential immersion of the samples in N-Methyl-2-pyrrolidone (NMP) and acetone baths followed by an isopropanol wash. 
\Yields{}

We employ AFM measurements to characterize the yield of our large-area nanodiamond assemblies.
The piezo-driven stage on the AFM has a maximum 40 $\mu$m $\times$ 40 $\mu$m scan area. 
We stitch together sequential scans to view the full arrays \cite{preibisch2009globally}.
Figure \ref{fig:Yields}A shows a representative AFM image of a TASA assembly of nanodiamonds in a 26$\times$26 array.
Although assembly yield is high (see below), we still observe instances of multi-particle and no-particle assembly. 
These imperfections are expected in part due to the polydispersity in size, shape, and surface chemistry of the nanodiamonds which can both impede assembly and lead to particle agglomeration. 
We also hypothesize that regional variations in assembly arise from fluctuations in chamber conditions, which alter the capillary forces, and incomplete liftoff, which may fail to remove larger nanodiamond deposits.
Dividing the larger array (Figure \ref{fig:Yields}A) into subarrays, as shown in Figure~\ref{fig:Yields}B, accounts for these regional variations and allows for statistical characterization of the assembly yield.

We repeat the assembly and AFM characterization for five different template trap diameters (Supporting Information Figure~S2). 
Collectively, we observe a median yield of single nanodiamonds of 76$\%$, with individual regions in most arrays reaching 100$\%$ yield (Figure~\ref{fig:Yields}B).
Figure~\ref{fig:Yields}B represents a lower bound on assembly yield, since higher resolution AFM scans occasionally reveal previously unseen  nanodiamonds.
No significant statistical variation in yield is observed for trap diameters ranging from 38--48 nm, however we see a $\sim$15$\%$ drop in yield for the 35~nm trap diameter template compared to those for the larger traps. 

\subsection{Nanodiamond Size Distribution}

In addition to detecting the presence of assembled nanodiamonds for yield estimates, AFM measurements quantify the nanodiamond heights. 
Tip convolution prevents reliable measurement of the lateral dimensions of individual nanodiamonds.
The distribution of particle heights (histogram, Figure~\ref{fig:Yields}D), with average size and standard deviation of 39 $\pm$ \SI{12}{\nano\meter}, is consistent with dynamic light scattering (DLS) measurements (red curve, Figure~\ref{fig:Yields}D; 36 $\pm$ \SI{12}{\nano\meter}) and with the manufacturer's specifications \cite{shenderova2017commercial, https://www.adamasnano.com/}.
Although the nanodiamonds are irregular in shape, 
we posit that the AFM height serves as a suitable measure for individual nanodiamond size.

Each template design yields arrays of nanodiamonds with a similar median size (Figure~\ref{fig:Yields}E).
The smallest \SI{35}{\nano\meter} template displays a tighter particle size range, while the larger templates have extended tails, especially towards larger particle sizes.
The observation of particles larger in height than the trap diameter is consistent with the likelihood of irregularly-shaped nanodiamonds assembling with their long axis perpendicular to the substrate surface.

\subsection{Optical Characterization\label{sec:OpticalChar}}


We use a custom-built, automated, confocal microscope (see Methods) to probe the optical and quantum properties of the nanodiamond arrays.
We stitch together multiple photoluminescence (PL) images to map the fluorescence of nanodiamonds in large arrays (Figure \ref{fig:Automation Overivews}A). 
Using an automated procedure, we characterize each fluorescent nanodiamond within a subarray (Figure~\ref{fig:Automation Overivews}B and Supporting Information Figure~S3~and~S4), recording measurements of the particles' PL spectrum, photon autocorrelation function, PL saturation as a function of excitation power, and spin lifetime.
Figures~\ref{fig:Automation Overivews}C-F depict the results of these measurements for four sites (I-IV) that exemplify the variation of observed optical properties. 
All optical measurements are performed using 532 nm excitation.
The excitation power is 0.425 mW, measured before the objective lens for all measurements except for saturation where the power varies.
AFM scans confirm that each of the four sites contains a single fluorescent nanodiamond (see Figure~\ref{fig:Yields}C and Supporting Information Figure~S3). 

\Measurements{}

Figure 3C shows single nanodiamond PL spectra.
Wavelengths below \SI{550}{\nano\meter} are cut off by the long pass filter used to exclude the \SI{532}{\nano\meter} excitation.
The NV center's PL spectrum consists of a linear combination of spectra associated with the NV$^{0}$ and NV$^{-}$ charge states. 
Under optical excitation, the NV center cycles between these two charge states through processes of ionization and recombination \cite{aslam2013photo}, resulting in different spectral weightings as shown by the examples of nanodiamonds I-IV.
We use a nonnegative matrix factorization method to decompose the spectra of individual nanodiamonds into their corresponding $NV^{0}$ (blue) and $NV^{-}$ (red) charge components \cite{reineck2019not}.
From these decomposed spectra we calculate the charge ratio,
$F_\mathrm{NV^{-}}
= S^{-}/(S^{-}+S^{0})$, where $S^{-}$ ($S^{0}$) is the integrated spectral intensity associated with the NV$^{-}$ (NV$^{0}$) charge state.
We graphically represent $F_\mathrm{NV^{-}}$ using a blue (predominantly NV$^{0}$) to red (predominantly NV$^{-}$) color scale throughout Figure~\ref{fig:Automation Overivews}.
Numeric results for nanodiamonds I-IV can be found in Table I. 
Observations of mixed charge states, $0<F_\mathrm{NV^{-}}<1$, can result from dynamic ionization and recombination of individual NV centers or from ensemble averaging over multiple NV centers within a nanodiamond.

 \begin{table*}[t]
 \centering
    \caption{Nanodiamond characteristics         \label{table:metrics}}
	\begin{tabular}{l  c c c c c c c }
        \hline
        \hline
	     \rule{0pt}{4ex} 
	     &\multicolumn{4}{c}{\textbf{Individual Nanodiamonds}} && \multicolumn{2}{c}{\textbf{Aggregate Results}} \\ 
       
	     & I & II & III & IV && Mean & Standard\\
	     & & & & & && Deviation \\ \hline
	    Height (nm) &21.4&32.9&41.5&41.7&&38.7 $\pm$ 0.1$^a$ &12.4$^a$\\ \hline

	   ${N}_\mathrm{Eff}$ &0.8$\pm$0.2 &2.2$\pm$0.2&4.1$\pm$0.4&9.3$\pm$2.6&&0.5$\pm$0.1$^a$&1.2$^a$  \\
	    &&&&&&2.1$\pm$ 0.1 &1.4\\
	   
	   \hline
	   	    $F_\mathrm{NV^{-}}$ &0.58&0.25&0.48&0.63&&0.61$\pm$ 0.1 &0.16 \\ \hline
	   ${C}_\mathrm{Sat}$ (kCts/s)  &14.0$\pm$0.5&77.0$\pm$3.1 &111.0$\pm$2.8&106.0$\pm$2.0&&28.6$\pm$ 0.1 &31.0 \\ \hline
	   $P_\mathrm{Sat}$ (mW) &0.12 $\pm$0.02 &0.25 $\pm$ 0.06&0.26$\pm$  0.04&0.13$\pm$ 0.01& &0.29$\pm$ 0.01 &0.42 \\ \hline
	   ${T}_1$ (\SI{}{\micro\second})&1.7$\pm$1.0&NA&125.4$\pm$55.6&65.0$\pm$14.7 &&11.58 $\pm$ 0.35 &24.5 \\ 
	   \hline
	   \hline\\
	  
        \end{tabular}\\
$^a$Including non-fluorescent nanodiamonds.
\end{table*}

We measure the photon autocorrelation function, $g^{(2)}(\tau)$, which 
 probes the likelihood of two fluorescence photons being detected at varying temporal delays, $\tau$ \cite{twiss1957correlation, fishman2021photon}.
Figure~\ref{fig:Automation Overivews}D shows examples of $g^{(2)}(\tau)$ for individual nanodiamonds.
$g^{(2)}(\tau)$ is corrected for background fluorescence, detector dark counts, and detector timing jitter. The zero-delay value, $g^{(2)}(0)$, is used to quantify the number of NV centers in each nanodiamond.
Specifically, we calculate the effective number (${N}_\mathrm{Eff}$) of NV centers  using the equation:
 \begin{equation}\label{eq:Neff}
 N_\mathrm{Eff}= \frac{1}{1- g^{(2)}(0)}
 \end{equation}
The progressively shallower dips seen in nanodiamonds I-IV for small delays, where $g^{(2)}(\tau)<1$ as $\tau \rightarrow 0$, indicate quantum fluorescence from a progressively larger number of discrete emitters. 
Nanodiamond I exhibits $g^{(2)}(0)<0.5$ which indicates that the emission is dominated by a single NV center.

This relationship assumes identical PL brightness from each emitter, which is not strictly accurate for nanodiamonds given their morphological inhomogeneities, varying crystal orientations, and different charge state populations.
Furthermore, measurements of $g^{(2)}(0)$ can be biased by systematic errors such as incomplete background correction, particularly for dim emitters (typical of small ${N}_\mathrm{Eff}$, \textit{e.g.}, for nanodiamond I) or when $g^{(2)}(0)\rightarrow 1$ (as ${N}_\mathrm{Eff}\rightarrow \infty$, \textit{e.g.}, for nanodiamond IV).
As described further in the Statistical Characterization Section, we find that ${N}_\mathrm{Eff}$ is linearly related to the saturation PL brightness of the nanodiamonds (Figure~\ref{fig:Automation Overivews}E), especially for the range $1<N_\mathrm{Eff}<10$ and we therefore interpret ${N}_\mathrm{Eff}$ as a reasonable quantitative estimate for the number of NV centers at each location.


While each nanodiamond varies in brightness (Figure 3B), by varying the illumination power, $P$, and recording the PL intensity, $I_\mathrm{PL}$, we find that the nanodiamond $I_\mathrm{PL}$ displays a similar characteristic power dependence (Figure 3E). 
As expected, the observed brightness  increases with $g^{(2)}(0)$ and the number of emitters, ${N}_\mathrm{Eff}$.
We perform this PL saturation measurement multiple times to check for stability, subtract a background measurement from a location on the sample nearby each nanodiamond, and fit the resulting data to the empirical model,
\begin{equation}
 \textrm{$I_\mathrm{PL}(P)$} = \frac{C_\mathrm{Sat}}{1 + \frac{P_\mathrm{Sat}} P},
 \end{equation}
from which we quantify the nanodiamond's saturation PL brightness  ($C_\mathrm{Sat}$) and saturation power ($P_\mathrm{Sat}$).

Finally, to probe the quantum features of the nanodiamonds, we measure the electron spin lifetime, ${T}_1$ (Figure 3F), using a time-resolved fluorescence technique.
We initialize the NV centers into their $m_s= 0$ spin-triplet sublevel using a \SI{1.6}{\micro\second},  \SI{532}{\nano\meter} laser excitation pulse and then count PL photons during a 300 ns readout laser pulse at delay times ranging from \SI{50}{\nano\second} to \SI{900}{\micro\second}. 
Due to the NV center's spin-dependent optical dynamics, the PL amplitude represents the probability that the NV centers remain in the $m_s=0$ spin sublevel of the $NV^{-}$ ground state at the time of the readout pulse. 
Longer ${T}_1$ lifetimes indicate higher levels of quantum isolation and enable improved quantum sensitivity \cite{tetienne2013spin, schirhagl2014nitrogen}. 
Variations at short ($\lesssim$200~ns) times can result from relaxation of the metastable spin singlet, whereas longer-time variations can reflect charge or spin relaxation \cite{ tetienne2013spin, schirhagl2014nitrogen,hopper2018amplified}.
We fit the data using a set of mathematical functions  that account empirically for the rates of these disparate processes as well as for the potential presence of multiple NV centers within a nanodiamond (see Methods). 
In some cases, such as nanodiamond II, the analysis is inconclusive; in these cases we do not report a ${T}_1$ lifetime.
 Nanodiamond II's small PL contrast is consistent with its low $F_\mathrm{NV^{-}}$, since spin relaxation is only observed for NV centers in the negative charge state.

\subsection{Statistical Characterization \label{sec:Statistics}}
	
Ordered arrays of nanodiamonds significantly simplify experiments, allowing us to study the statistical properties of large numbers of nanodiamonds, correlating optical and structural measurement modalities.
We collect AFM scans and the suite of optical measurements, as described for the example nanodiamonds in Figure 3, but now for 219 nanodiamonds assembled in multiple arrays, to build statistical data sets of nanodiamond size and emitter number, brightness, charge state, and spin lifetime (Figure~\ref{fig:Data Overivews}).
The Supporting Information also includes an analysis of correlations between these variables (Supporting Information Figures~S5--S9). 

\Histograms{}

\subsubsection{Emitter Number Distribution}
We study the number of NV centers hosted within each nanodiamod (Figure ~\ref{fig:Data Overivews}A). 
A significant fraction (68 out of 219, or 31\%) of nanodiamonds are observed in AFM measurements but invisible in optical measurements.
We set ${N}_\mathrm{Eff}=0$ for these non-fluorescent nanodiamonds, as indicated by the purple bar in Figure~\ref{fig:Data Overivews}A. 
The non-fluorescent nanodiamonds are excluded from the other datasets in Figure ~\ref{fig:Data Overivews}.
For the fluorescent nanodiamonds, we calculate ${N}_\mathrm{Eff}$ from autocorrelation measurements of $g^{(2)}(0)$, as in Figure~3D, using eq.~(\ref{eq:Neff}). 
Throughout Figure ~\ref{fig:Data Overivews}, we indicate the nanodiamonds likely to contain one ($0<N_\mathrm{Eff}<1.5$) or two ($1.5< N_\mathrm{Eff}<2.5$) NV centers using separately colored bars.
Of the 151 fluorescent nanodiamonds, four are too dim and sixteen are too bright to reliably determine ${N}_\mathrm{Eff}$ from $g^{(2)}(0)$;
we estimate ${N}_\mathrm{Eff}$ for these nanodiamonds using the observed linear relationship between ${N}_\mathrm{Eff}$ and ${C}_\mathrm{Sat}$, as described below  (Figure \ref{fig:Data Overivews}B, inset) and add the results to Figure~\ref{fig:Data Overivews}A as grey bars. 

The observed fractions of non-fluorescent nanodiamonds (31\%) and single emitters (12\%) in this sample are in close agreement with the manufacturer's product specifications ($25\%$ and $13\%$, respectively)\cite{https://www.adamasnano.com/}. 
However, we show below that the observed $N_\mathrm{Eff}$ distribution in Figure~\ref{fig:Data Overivews}A is not consistent with a simple stochastic model assuming an isotropic distribution of NV centers within the nanodiamonds.
We develop a quantitative model for the observed distribution, motivated by the mechanisms of diamond growth, irradiation, and milling.

A model for the ${N}_\mathrm{Eff}$ distribution must account for the variance in particle size and the stochastic incorporation of NV centers within each particle.
First, we consider a model in which the probability of observing $N$ emitters in a spherical nanodiamond of diameter $d$ obeys a Poissonian distribution, $P(N|d)=\mathrm{Poiss}(N;\langle N\rangle)$, where the mean, $\langle N\rangle=\rho V$, is determined by the NV density, $\rho$, and the particle volume, $V=\pi d^3/6$.  
The probability distribution for $N$ is subsequently given by $\sum_{i}P(N|d_{i})\phi(d_{i})$, where $\phi(d)$ is the probability of finding a particle with diameter $d$, as characterized by AFM measurements (Figure~\ref{fig:Yields}D).
This model fails to match the observed distribution, with a marked discrepancy especially at low ${N}_\mathrm{Eff}$. (The best fit is rejected on statistical grounds with reduced chi-squared $(\chi_r^2)=3.05$, $p < 10^{-10}$, as shown in Supporting Information Figure~S10.)
However, when we exclude the non-fluorescent nanodiamonds, this model can explain the distribution of remaining fluorescent particles ($\chi_r^2=1.2$; $p=0.29$). 
We conclude that the relatively large fraction of non-fluorescent particles in our data cannot be explained using this simple model.

To reconcile this discrepancy, we hypothesize that the nanodiamonds can be divided into two groups: one where the NV centers are distributed isotropically with non-negligible density, resulting in the observed ${N}_\mathrm{Eff}$ distribution of fluorescent nanodiamonds, and another group where the density is so low that there is a negligible probability of finding an NV center.
Nitrogen is incorporated inhomogeneously during high-pressure, high-temperature growth of bulk diamond crystals \cite{shenderova2019synthesis, babich2009spatial, kanda1993inhomogeneous}.
When the crystals are milled, the resulting nanodiamonds have an inhomogeneous nitrogen concentration. 
Subsequent electron irradiation creates a homogeneous distribution of carbon vacancies, which become mobile upon annealing to form NV centers. 
In regions with high nitrogen content, the NV-center formation will be vacancy limited and hence the NV-center density is proportional to the radiation dose. 
In nitrogen-depleted regions, the NV-center formation will be nitrogen limited, and the density will be lower.
This hypothesis is supported by the observation that Ad\'amas Nanotechnology produces ``highly-fluorescent'' nanodiamonds with the same size distribution, using an identical synthesis and milling procedure but significantly higher irradiation levels. These samples also contain a 30\% fraction of non-fluorescent particles.
Furthermore, the height distribution of non-fluorescent particles (Figure ~\ref{fig:Data Overivews}A, inset) is similar to the full distribution (Figure ~\ref{fig:Yields}D), whereas an exclusively isotropic NV-center creation model would predict a narrowed distribution of primarily non-fluorescent smaller particles. 
The invariance of the non-fluorescent fraction with respect to milling and irradiation conditions supports the conclusion that the heterogeneity in NV-center creation derives from a nonuniform nitrogen distribution. 

To account for this effect in our model, we adopt a bimodal distribution of NV-center density, where $\rho=\rho_1$ with probability $\alpha$, and $\rho=\rho_2$ with probability $1-\alpha$; see Methods, Equation (\ref{eq:Bimodal}).
A fit to the data using this model yields an acceptable result as shown by the curve in Figure ~\ref{fig:Data Overivews}A ($\chi_r^2=1.3$, $p=0.17$), where $\rho_1 =  14 \pm 1.4 \times 10^{16}$ cm$^{-3}$, $\rho_2 =  0.001 \pm 0.45 \times 10^{16}$ cm$^{-3}$, and $\alpha =  0.77 \pm 0.05$.
The implication of this analysis is that the observed 31\% of dark nanodiamonds arise from a combination of the underlying material inhomogeneity (accounting for 23\% of dark nanodiamonds in the sample) and the stochastic creation of NV centers (accounting for the remaining 8\% of dark nanodiamonds).
Accounting for the expected ranges of NV-center creation efficiency, these fits place lower (upper) limits on the nitrogen content in the nitrogen-rich (nitrogen-poor) particles of 7.1 ppm (2.5 ppm).
See the Supporting Information section III for more details on the nitrogen content and NV-center creation estimates.

\subsubsection{Nanodiamond Brightness}

The empirical saturation brightness distribution is shown in Figure~\ref{fig:Data Overivews}B.
We observe a clear linear relationship between ${C}_\mathrm{Sat}$ and ${N}_\mathrm{Eff}$ (Figure \ref{fig:Data Overivews}B inset), with a slope of 22.8 $\pm$ 2.2 kCts/s per NV center.
The linear relationship implies that NV centers in these nanodiamonds have nearly identical apparent brightness, despite wide variations in charge stability and spin stability, as discussed in the following sections.
The strong linear relationship between ${C}_\mathrm{Sat}$ and ${N}_\mathrm{Eff}$ implies that more time-efficient measurements of ${C}_\mathrm{Sat}$ in comparison to $g^{(2)}(0)$ can serve as a robust indicator for ${N}_\mathrm{Eff}$. 
We use the relationship to estimate ${N}_\mathrm{Eff}$ values for the 20 nanodiamonds with an ambiguous  $g^{(2)}(0)$ in our dataset (Figure~\ref{fig:Data Overivews}, grey bars and Supporting Information Figure~S11--S12). 
The distribution for ${P}_\mathrm{Sat}$ can be found in Supporting Information Figure~S13.

\subsubsection{NV-Center Charge-State Distribution}

Figure~\ref{fig:Data Overivews}C shows the $F_\mathrm{NV^{-}}$ distribution for fluorescent nanodiamonds in our assemblies.
The mean value ($\langle F_\mathrm{NV^{-}}\rangle=0.6\pm0.1$; see Table~\ref{table:metrics}) agrees with previous measurements of randomly dispersed, $\sim$150-nm-diameter, milled nanodiamonds \cite{reineck2019not}.
However, it is significantly lower than the measurement of $F_\mathrm{NV^{-}} =$ 0.75 for the parent nanodiamond dispersion used in assembly (Supporting Information Figure~S14).
The dispersion and individual nanoparticle measurements are performed at similar optical excitation power densities, however we note that dispersed particles can exchange freely in and out of the excitation path during the measurement, whereas assembled, individual particles are continuously probed. 
Although ionization and recombination rates depend on excitation power density, the ratio $F_\mathrm{NV^{-}}$ generally does not vary strongly with power under 532 nm excitation \cite{aslam2013photo, hacquebard2018charge, waldermann2007creating}.
We did not observe a power dependence in the qualitative shape of the PL spectra of individual nanodiamonds for excitation powers ranging from  $0.1$ to $1.6$  mW before the objective.

Accordingly, the nanoparticles' environment may play the dominant role in determining $F_\mathrm{NV^{-}}$.
A bias towards the negative charge state is often explained by a combination of core nitrogen impurities, which are reported to donate electrons to NV centers \cite{collins2002fermi, waldermann2007creating}, together with the presence of carboxylate anions on the nanodiamond surface \cite{rondin2010surface, reineck2019not, petrakova2015charge}. 
Once assembled, we hypothesize that silanol groups or water of hydration on the SiO$_{2}$ surface, known to act as electron traps \cite{aguirre2009role, robertson1984intrinsic, shelby1977molecular}, can neutralize the negative surface charge of the nanodiamonds and thereby decrease $F_\mathrm{NV^{-}}$.

A closer examination of the $F_\mathrm{NV^{-}}$ distribution of individual nanodiamonds containing small numbers of NV centers reveals a bimodal shape (Figure~\ref{fig:Data Overivews}C).
Nanodiamonds with one or two NV centers exhibit $F_\mathrm{NV^{-}}$ values clustered either around 0.4 or around 0.9. 
A few particles exhibit $F_\mathrm{NV^{-}}>0.95$, essentially a pure negative charge-state configuration.
The bimodal distribution for nanodiamonds with ${N}_\mathrm{Eff}\sim$1-2 suggests the NV centers fall into two distinct categories based on $F_\mathrm{NV^{-}}$.
The more mixed charge states ($F_\mathrm{NV^{-}}$ $\sim 0.6$) seen in particles with larger ${N}_\mathrm{Eff}$ therefore arise from averaging the signals from multiple NV centers. 
The  bimodal distribution could reflect distinct local environments, for example, distinguishing NV centers near the surface from those within the nanodiamond core.
We checked for a correlation between $F_\mathrm{NV^{-}}$ and particle size, which would support this interpretation, but any such correlation is not significant in our data (see Supporting Information Figures~S9-10). 
However, if environmental effects are limited to NV centers within a few nanometers from a particle's surface, we would not expect to observe a strong correlation for this range of particles sizes. 
To quantify this relationship, we compared the fraction of nanodiamonds with $F_\mathrm{NV^{-}}<0.5$ with the predicted shell-volume fraction as a function of particle size (see Supporting Information Figure~S15).
We find that our data are consistent with this interpretation, where NV centers located within 2-3~nm of the surface have reduced $F_\mathrm{NV^{-}}$, however even larger sample sizes would be required to establish a direct correlation between $F_\mathrm{NV^{-}}$ and particle size.

The observation of a bimodal charge-state distribution has important practical implications.
The pure negative charge state is typically desired for quantum spin sensing applications, whereas mixed charge states may be more sensitive to variations in electrochemical potential \cite{karaveli2016modulation}.
Hence, deeper understanding of this phenomena that leads to manufacturers' ability to promote one category over the other, or to efficiently sort the particles based on $F_\mathrm{NV^{-}}$, would improve the material's performance in these key applications.

\subsubsection{Spin Lifetimes}

Spin relaxation measurements are performed as described in the Optical Characterization section, and those nanodiamonds whose measurements are adequately fit by our models are included in Figure~\ref{fig:Data Overivews}D; we are unable to resolve ${T}_1$ values for 48 nanodiamonds. 
For the remaining nanodiamonds, we observe a wide distribution of ${T}_1$ values, ranging from $1.0 \pm 0.5  ~\mu$s to $608.0 \pm 211.5  ~\mu$s; see Figure~\ref{fig:Data Overivews}D.
We also observe a positive correlation between ${T}_1$ and $F_\mathrm{NV^{-}}$ (see Supporting Information Figures~S5 and S9). 
This correlation is consistent with the interpretation proposed in the previous section that NV centers close to the nanodiamond's surface display higher environmental sensitivity, as seen by lower $F_\mathrm{NV^{-}}$ and ${T}_1$, according to a core-shell relationship. 
The lack of a correlation with height, which has been observed for studies considering much larger size ranges\cite{tetienne2013spin, ong2017shape}, further stresses the importance of NV center placement rather than particle size for these nanodiamonds. 
However, other interpretations are  possible.
For example, proximal fluctuating charges that lower $F_\mathrm{NV^{-}}$ for an NV center in the nanodiamond core could also limit $T_1$.
Charge fluctuations have been observed for NV centers in nanodiamonds in the dark following a laser pulse \cite{hopper2018amplified}; such fluctuations could systematically bias the ${T}_1$ measurements.

As for the case of charge-state stability, understanding and controlling the spin lifetime has important implications for applications.
Particularly for spin relaxometry sensing and imaging, an increased ${T}_1$ translates directly to improved sensitivity \cite{rondin2014magnetometry,schirhagl2014nitrogen,tetienne2013spin,hopper2018amplified}.
We note that several of the longest ${T}_1$ values observed in our sample occurred in nanodimaonds with only one or two NV centers, boding well for the prospects of single-spin applications.

\section{Conclusion}
We describe the use of the TASA method to assemble large arrays of single nanodiamonds with high yield. 
We exploit these arrays to perform automated measurements of the morphological, optical, and quantum properties of nanodiamonds from 219 isolated particles.
We observe a wide distribution of emitters in the particles, ranging from zero to $>$20 NV centers per particle, with 31\% dark nanodiamonds and 12\% single NV centers. 
We attribute the distribution to spatial heterogeneity in the nitrogen incorporation during growth of the parent, bulk crystal as well as stochastic creation of NV centers within nanodiamonds.
Greater control over the nitrogen incorporation rates during crystal growth, vacancy generation during electron irradiation, and the particle size uniformity could provide tighter control over the NV center distribution.
The fluorescent nanodiamonds show a wide distribution in their charge states and spin lifetimes.
We find that $F_\mathrm{NV^{-}}$ and $T_{1}$ are not correlated with nanodiamond size in this sample, but instead indicate a heterogeneity in the NV-center environment, which could arise from variations in the internal crystal structure and composition or the NV center distance from the nanodiamond surface. 
Controlling the surface conditions should therefore enable more uniform charge states and longer spin lifetimes.

Building on the TASA method and leveraging our understating of the NV center's properties, we see a path for integrating nanodiamonds and other nanoscale quantum materials into photonic and electronic quantum devices. 
We envision extending the TASA method for the placement of individual nanodiamonds to that of other irregularly shaped nanoparticles.
Templates can also be fabricated on functional substrates to better optimize the nanodiamonds' surface environment.
We can adapt the TASA method to create multiparticle heterostructure assemblies \cite{greybush2014plasmon, bogdanov2017electron, bogdanov2019hybrid}, where plasmonic, magnetic, or dielectric nanoparticles could act as optical or magnetic antennas and enhance the NV-center's optical signal and quantum sensitivity.

\section{Methods}

\subsubsection{Template Fabrication and Nanodiamond Assembly}

The starting material for these studies is an aqueous suspension of 40 nm, milled, fluorescent nanodiamonds (Ad\'amas Nanotechnology, NDNV40nmLw10ml). 

We fabricate templates on single-side-polished Si wafers with a 250 nm SiO$_2$ thermal oxide layer (Virginia Semiconductor). 
Poly(methyl methacrylate) (PMMA) positive resist, 950 K molecular weight PMMA in anisole (MicroChem), is spun onto the substrates at 3000 rpm for 60 s followed by a soft bake at 180 $^o$C for 90 s. 
Electron-beam lithography (Elionix ELS-7500EX) is used to pattern the template at a 50 kV accelerating voltage and a 50 pA beam current.
Circular traps are designed with diameters ranging from 35--200 nm. 
The PMMA is developed in a 1:3 water:isopropanol solution for 90 s. 
Prior to assembly we perform  a 6 s, 75 W, 20 sccm, O$_2$ plasma (Gatan Solarus) descum resulting in an approximately 62-nm-thick template.

The assembly is performed using a custom-built capillary assembly apparatus \cite{greybush2014plasmon}.
A 35 $\mu$L volume of a 0.001 mg/mL nanodiamond dispersion in a 1\% Sodium dodecyl sulfate (Sigma-Aldrich)  aqueous dispersion is dropped between the template and a glass slide. 
The substrate is translated by a motorized linear stage (New Focus Picomotor) at a rate of 3.5 $\mu$m/s. The assembly apparatus is enclosed in a home-built, humidity-controlled environmental chamber, and the substrate temperature is regulated by a water-cooled stage and monitored by a thermocouple.
During assembly, the substrate temperature is 24  $^{\circ}$C, and the ambient dew point is maintained between 8.5 and 9.5  $^{\circ}$C. 
These temperature and humidity  conditions maintain a contact angle $>$24$^{\circ}$ throughout the assembly. 
Lift off is then performed by soaking the assembled samples in N-Methyl-2-pyrrolidone (NMP) for 1 min, followed by 5 min acetone and IPA washes.
\subsubsection{AFM  Measurements}

Atomic force microscopy (AFM) measurements are taken using a MFP-3D-BIO AFM from Asylum Research. All measurements are collected using a standard Si probe (AC240TS-R3, Olympus). 
For quantifying yields,
large-area scans (\SI{40}{\micro\meter} $\times$ \SI{40}{\micro\meter}) are taken at a scan rate of 0.3 Hz.
Mid-area scans (\SI{20}{\micro\meter} $\times$ \SI{20}{\micro\meter}) for mapping subarrays are taken at a scan rate of 0.27 Hz and individual nanodiamond measurements are collected using a \SI{1.5}{\micro\meter} $\times$ \SI{1.5}{\micro\meter} scan area at a scan rate of 0.3 Hz. Igor Pro version 6.38B01 is used to analyze AFM data. Composite images are formed from the AFM scans using ImageJ and the program described by \citet{preibisch2009globally}
Total yield counts and are analyzed by hand from the resulting images.

\subsubsection{Optical Measurements}
Optical measurements are taken using two custom-built confocal microscopes. 
In both microscopes a fast steering mirror (FSM, Optics in motion, OIM101) projects a beam onto to the back of a 100x objective (0.9 NA 100x objective Olympus, MPI Plan Fluor/OL, or Nikon, Plan Flour x100/0.5-1.3), allowing us to raster a 532~nm (green) continuous-wave laser (Coherent, Compass 315M-150) over the sample and collect PL emission on single-photon avalanche diode detectors.
We employ two microscopes to allow multiple measurements to run in parallel and to extend the spectral bandwidth. 
The first microscope, described by \citet{exarhos2019magnetic}, is used to collect saturation and autocorrelation measurements.
The  second microscope, described by \citet{huang2019monolithic}, is modified with a 532 nm low pass dichroic mirror (Semrock), to extend the spectral bandwidth and collect PL maps, spectra, and ${T}_1$ measurements. 
Throughout each measurement, a periodic tracking PL scan is used to maintain alignment for optimal collection from individual  nanodiamonds.
During tracking, PL line scans in the $x$ and $y$ directions (controlled by the FSM) are fit using a Gaussian peak plus a constant background. 
The FSM is then adjusted to align the microscope to the Gaussian's mean position. 
A similar tracking scan and Gaussian fit in the $z$ (focus) direction is also performed utilizing a piezoelectric stage (Thorlabs MZS500-E) connected to either the sample mount or the objective, depending on the microscope, to optimize focus.

Spectral measurements are taken on a Princeton Instruments IsoPlane-160 spectrometer with a 700-nm blaze, 300 G/mm grating and a thermoelectrically cooled charge-coupled device (CCD) camera (Princeton Instruments PIXIS
100BX) yielding a spectral resolution of 0.7 nm. 
Measurements consist of one 90 s background measurement combined with multiple  90 s signal measurements. 
We then perform  cosmic ray rejection and background subtraction on each scan before combining the signals and correcting for wavelength-dependent photon detection efficiencies.

The power-dependent PL measurements  are achieved by placing a variable neutral density filter (Thorlabs NDC-50C-2-A) in the excitation path, with pre-objective powers calibrated with a Thorlabs PM100D power meter.

$T_1$ lifetime measurements are implemented by programming the required pulse sequences (described in the Optical Characterization subsection) onto an arbitrary-waveform generator (AWG; AWG520 Tektronix).  
The AWG control signals are passed to the microscope's  acousto-optic modulator (AOM) for generating optical pulses, and to three high isolation switches (ZASWA-250DR Mini-Circuits) for time-gating photon detection events recorded by counters in a data acquisition card (National Instruments PCIe-6323). 

DLS measurements are conducted on a Malvern Instruments Zetasizer  Nano-s and analyzed using Malvern's software. 

\subsubsection{Autocorrelation analysis}

Autocorrelation data is collected using a Hanbury
Brown-Twiss setup with a time correlated single-photon counting module (PicoQuant PicoHarp 300) in time-tagged, time-resolved collection mode.
Background correction is performed by determining $\rho =S/(S+B)$ from the signal ($S$) and background ($B$) levels for each spot (both determined from Gaussian fits during tracking scans) and then calculating the corrected autocorrelation function:
\begin{equation}
     g_\mathrm{corr}^{(2)}(\tau) =  \frac{g_\mathrm{meas}^{(2)}(\tau) -(1 - \rho^2)}{\rho^2}
 \end{equation}
The background corrected autocorrelation data are then fit using the empirical model:
\begin{equation}
g_\mathrm{corr}^{(2)}(\tau) = 1 + A - De^{-|\tau-t_{0}|/\tau_{1}}
\end{equation}
Here, $A$ and $D$ represent the bunching and antibunching amplitudes, respectively, $t_{0}$ accounts for signal delays, and $\tau_{1}$ represents the antibunching timescale
\cite{fishman2021photon}.

The signal-to-noise ratio of autocorrelation measurements depends on $S$, $\rho$, and on the acquisition time. 
Therefore, we dynamically adjust the acquisition time for each nanodiamond in order to achieve a desired uncertainty in $g^{(2)}(\tau)$ ($\Delta g$), given a particular time resolution ($\Delta \tau$): 
  \begin{equation}
  \textrm{Measurement Time } =  4 \frac{(S + B)^2} { S^4 (\Delta g)^2 \Delta \tau }
 \end{equation}
 
\noindent For this study, we use $\Delta\tau=2$~ns, and we set $\Delta g=0.025$ in order to achieve sufficiently high resolution, however we limit the total measurement time for individual spots to a maximum of 1 h. 
For most instances, this results in a resolvable $g^{(2)}(0)$ value. 
For some exceptionally bright or dim nanodiamonds, the antibunching dip
is smaller than $\Delta g$, and we cannot reliably determine ${N}_\mathrm{Eff}$ values from $g^{(2)}(\tau)$.

\subsubsection{Fitting Spin Relaxation Data}

Recorded spin lifetime measurements for single NV centers are typically fit to a four-level rate model \cite{tetienne2013spin}:
\begin{equation}
  I(\tau)  = I(\infty)  [1 - C_{m}e^{\frac{-\tau}{T_m}} + C_{1}e^{\frac{-\tau}{T_1}}]
 \end{equation}
where $T_m$ accounts for the relaxation of the metastable singlet spin state.
To account for the presence of multiple NV centers, we expand our empirical fitting model to account for a second spin relaxation rate:

 \begin{equation}
  I(\tau)  = I(\infty)  [1 - C_{m}e^{\frac{-\tau}{T_m}} + C_{1}e^{\frac{-\tau}{T_1}} + C_{2}e^{\frac{-\tau}{T_2}}]
 \end{equation}

\noindent Due to signal-to-noise constraints and the potential overlap with $T_1$, we may not always observe the $T_m$ term. 
Accordingly, we add two more potential models to describe the spin relaxation dynamics of the nanodiamonds. 

 \begin{equation}
  I(\tau)  = I(\infty)  [1 + C_{1}e^{\frac{-\tau}{T_1}}]
 \end{equation}
 
 \begin{equation}
  I(\tau)  = I(\infty)  [1  + C_{1}e^{\frac{-\tau}{T_1}} + C_{2}e^{\frac{-\tau}{T_2}}]
 \end{equation}

\noindent We fit the data for each nanodiamond using all four models. 
We exclude fits where uncertainties for the  $T_{1}$ parameter exceed the $T_{1}$ result, where parameters are inconsistent with known physical properties (${T_m} > 10$ $\mu$s or $C_{1} < 0$), or where $\chi_r^2$ is greater than 1 $ + 2 \sqrt{\frac{2}{ \mathrm{DoF}}}$.
Here  $\mathrm{DoF}$ is the number of degrees of freedom in the model.
If no valid fits remain, we exclude that data point from later statistical analysis. 
In the event of multiple acceptable fits, $\chi_r^2$ is within  1 $\pm \sqrt{\frac{2}{\mathrm{DoF}}}$, we choose the model with the smallest number of parameters.

\subsubsection{Modeling the Emitter Number Distribution}
 We fit the $N_\mathrm{Eff}$ data by modeling the total probability of observing $N$ emitters given a spherical nanodiamond of diameter $d$ weighted by the likelihood of observing each $d$ value:
 
 \begin{equation}\label{eq:Possion}
  P(N)  = \sum_{i} P(N | d_{i}) \phi(d_{i})
 \end{equation}
Here, the conditional  probability of $N$ given $d$ is modeled as a Poissonian distribution: $P(N|d)=\mathrm{Poiss}(N;\langle N\rangle)$, where the mean, $\langle N\rangle=\rho V$, is determined by the NV density, $\rho$, and the particle volume, $V=\pi d^3/6$. 
The unconditional probability of finding a particle with diameter $d$, $\phi(d)$, is determined using AFM measurements as shown in Figure~\ref{fig:Yields}D. 
To account for a variation in nitrogen content, we assume two different $\rho$ values with their own probabilities $\alpha$ and $ 1 - \alpha$. 
The model of P($N$) then becomes:

  \begin{equation}\label{eq:Bimodal}
  P(N)= 
  \sum_{i}\left[ 
  P(N |d_{i}, \rho_{1}) \phi(d_{i})
  \alpha
  + P(N |d_{i}, \rho_{2}) \phi(d_{i})
  (1-\alpha) \right] 
 \end{equation}
The results of these models can be found in Supporting Information Figure S10.

\begin{acknowledgements}
This research was primarily supported by NSF through the University of Pennsylvania Materials Research Science and Engineering Center (MRSEC) (DMR-1720530). 
The authors thank Yun Chang Choi for generating the 3D schematic of the  nanodiamond assembly process.

\end{acknowledgements}


\bibliography{Ref}
\end{document}



\title{ Supporting Information For: \\
Template-Assisted Self Assembly of Fluorescent Nanodiamonds for Scalable Quantum Technologies}

\author{Henry J. Shulevitz}
\affiliation{Department of Electrical and Systems Engineering \\
University of Pennsylvania, Philadelphia PA, 19104, USA}
\author{Tzu-Yung Huang}
\affiliation{Department of Electrical and Systems Engineering \\
University of Pennsylvania, Philadelphia PA, 19104, USA}
\author{Jun Xu}
\affiliation{Department of Electrical and Systems Engineering \\
University of Pennsylvania, Philadelphia PA, 19104, USA}
\author{Steven Neuhaus}
\affiliation{Department of Materials Science and Engineering\\
University of Pennsylvania, Philadelphia PA, 19104, USA}
\author{Raj N. Patel}
\affiliation{Department of Electrical and Systems Engineering \\
University of Pennsylvania, Philadelphia PA, 19104, USA}
\author{Lee C. Bassett}
\email[Corresponding authors. ]{lbassett@seas.upenn.edu \& kagan@seas.upenn.edu }
\affiliation{Department of Electrical and Systems Engineering \\
University of Pennsylvania, Philadelphia PA, 19104, USA}
\author{Cherie R. Kagan}
\email[Corresponding authors. ]{lbassett@seas.upenn.edu \& kagan@seas.upenn.edu }
\affiliation{Department of Electrical and Systems Engineering \\
University of Pennsylvania, Philadelphia PA, 19104, USA}
\affiliation{Department of Materials Science and Engineering\\
University of Pennsylvania, Philadelphia PA, 19104, USA}
\affiliation{Department of Chemistry\\
University of Pennsylvania, Philadelphia PA, 19104, USA}
  
\date{\today}
             

\def \Solution{\begin{figure}[H]
\renewcommand\figurename{Supporting Information Figure}
\includegraphics[width = \textwidth ]{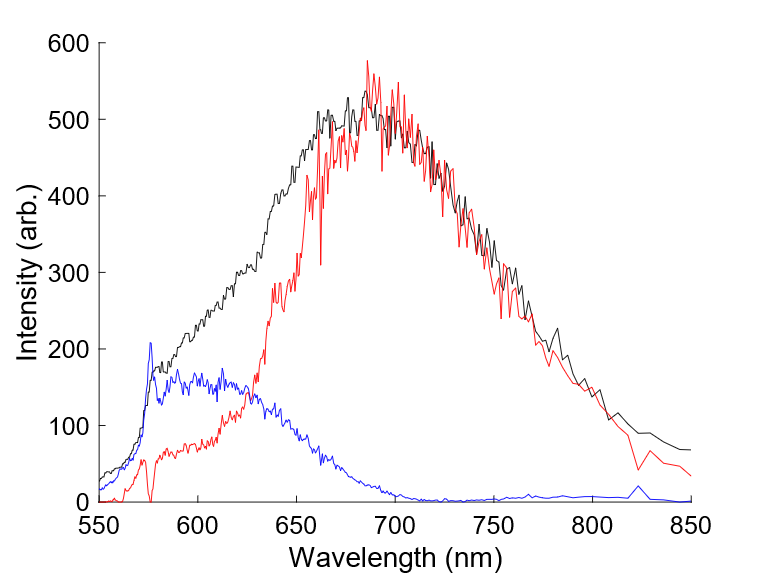}
\caption{\label{fig:DispersionSpectra}
 Spectral measurement (black) and nonnegative matrix factorization (red $NV^{-}$, blue $NV^{0}$) decomposition for a 1 mg/mL dispersion of nanodiamonds.
 F$_{NV^{-}}$ = .75.
}

\end{figure}
}

\def \CSatVN{\begin{figure}[H]
\renewcommand\figurename{Supporting Information Figure}
\includegraphics[width = \textwidth ]{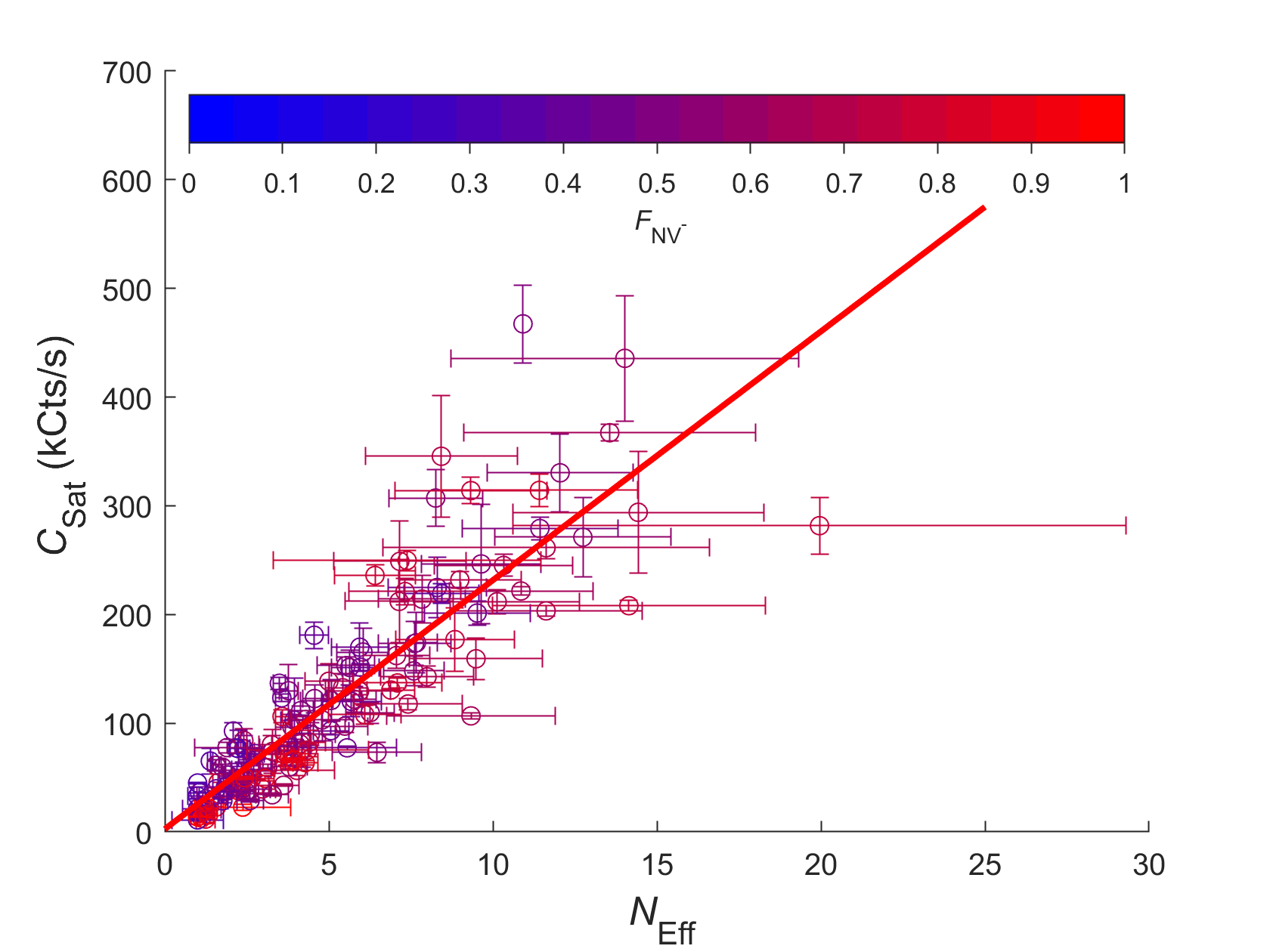}
\caption{\label{fig: CSat Vs N}
Plot of  $C_\mathrm{Sat}$ \textit{vs} $N_\mathrm{Eff}$ for 131 nanodiamonds with reliable  $N_\mathrm{Eff}$  values, with linear fit (red). 
}

\end{figure}}
\def \Corrlations{\begin{figure}[H]
\renewcommand\figurename{Supporting Information Figure}
\includegraphics[width = \textwidth ]{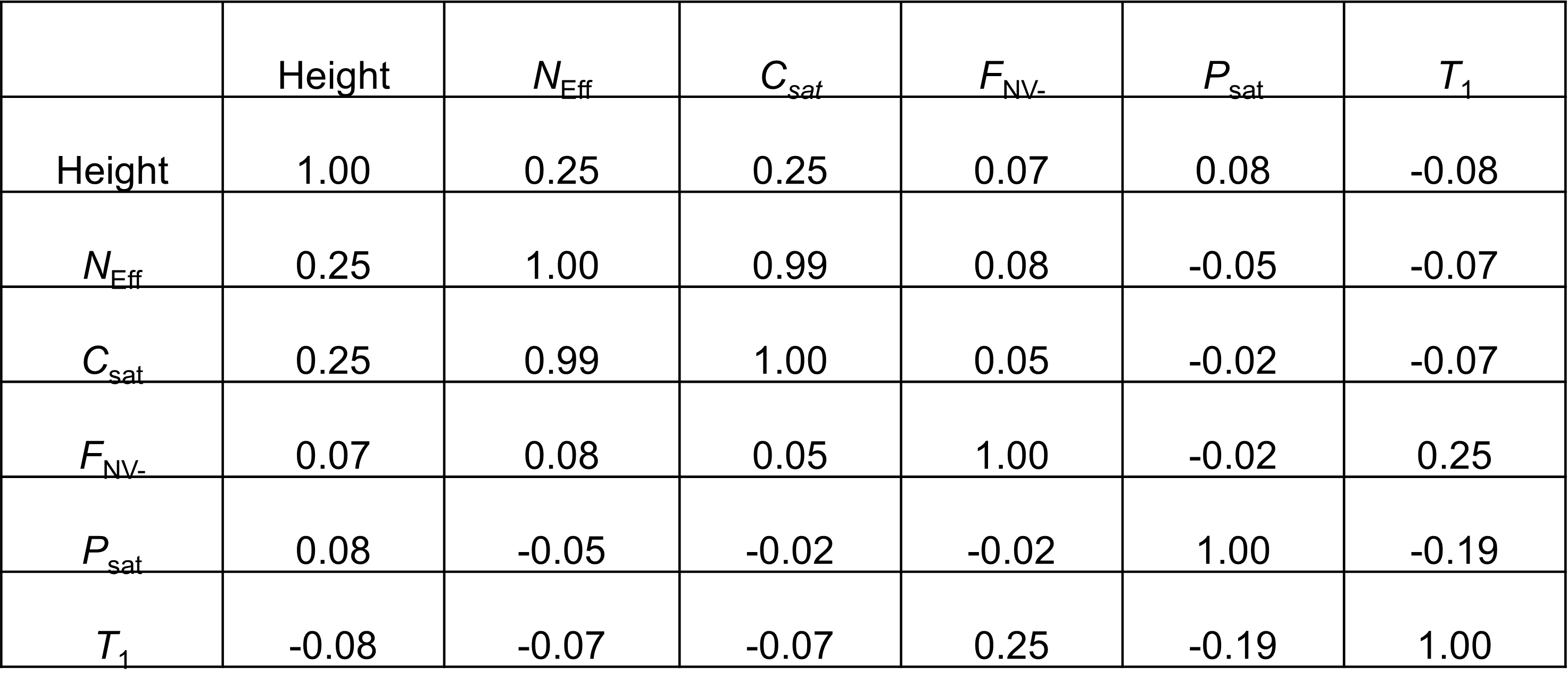}
\centering
\caption{\label{fig: Correlations}
Pearson correlation coefficients for different optical and morphological properties of the nanodiamonds. 
A  magnitude $\geq 0.2$ indicates a potentially significant correlation. 
}

\end{figure}}

\def \CorrlationsPVal{\begin{figure}[H]
\renewcommand\figurename{Supporting Information Figure}
\includegraphics[width = \textwidth ]{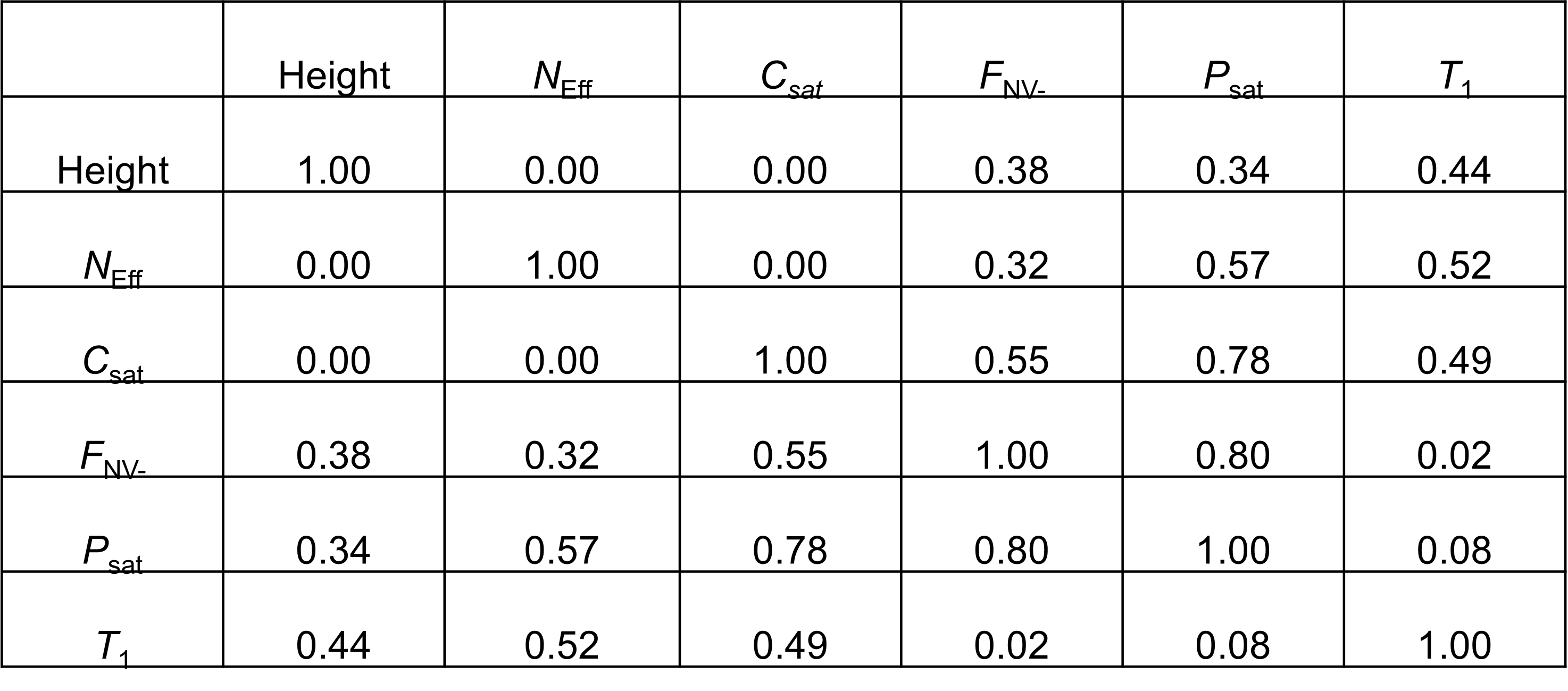}
\caption{\label{fig: Correlations_PVal}
Pearson correlation coefficients $p$-Values for different optical and morphological properties of the nanodiamonds. 
A  $p$-Value $\leq 0.05$ indicates a potentially significant correlation. 
}

\end{figure}
}

\maketitle

\def \CoreShell{\begin{figure}[H]
\centering
\renewcommand\figurename{Supporting Information Figure}
\includegraphics[scale=.85]{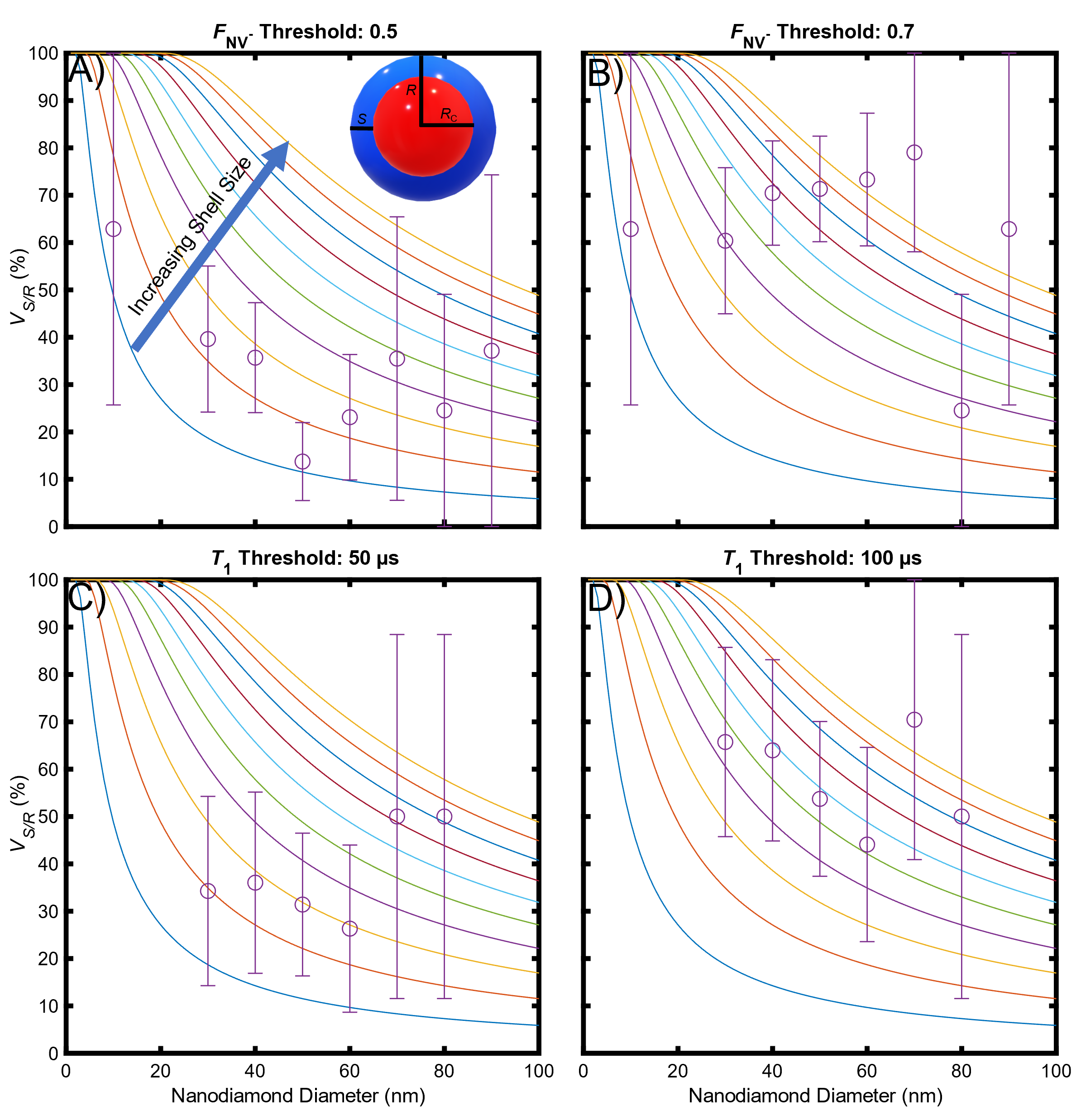}
\centering
\caption{\label{CoreShell} Throughout the figure, colored lines indicate the shell volume ratio ($V_{{S}/{R}}$) for shell sizes ranging from 1-10~nm. 
In each panel, circles and error bars represent the Wilson score interval estimate for $V_{{S}/{R}}$ based on the number of $F_\mathrm{NV^-}$ (A,B) or $T_1$ measurements (C,D) that fall below the threshold listed in each plot label, for nanodiamonds binned according to their height.
The inset of (A) shows a core-shell structure with total radius $R$, core radius $R_{C}$ and shell thickness $S$. 
}

\end{figure}
}

\def \PofNFits{\begin{figure}[H]
\renewcommand\figurename{Supporting Information Figure}
\includegraphics[width = \textwidth ]{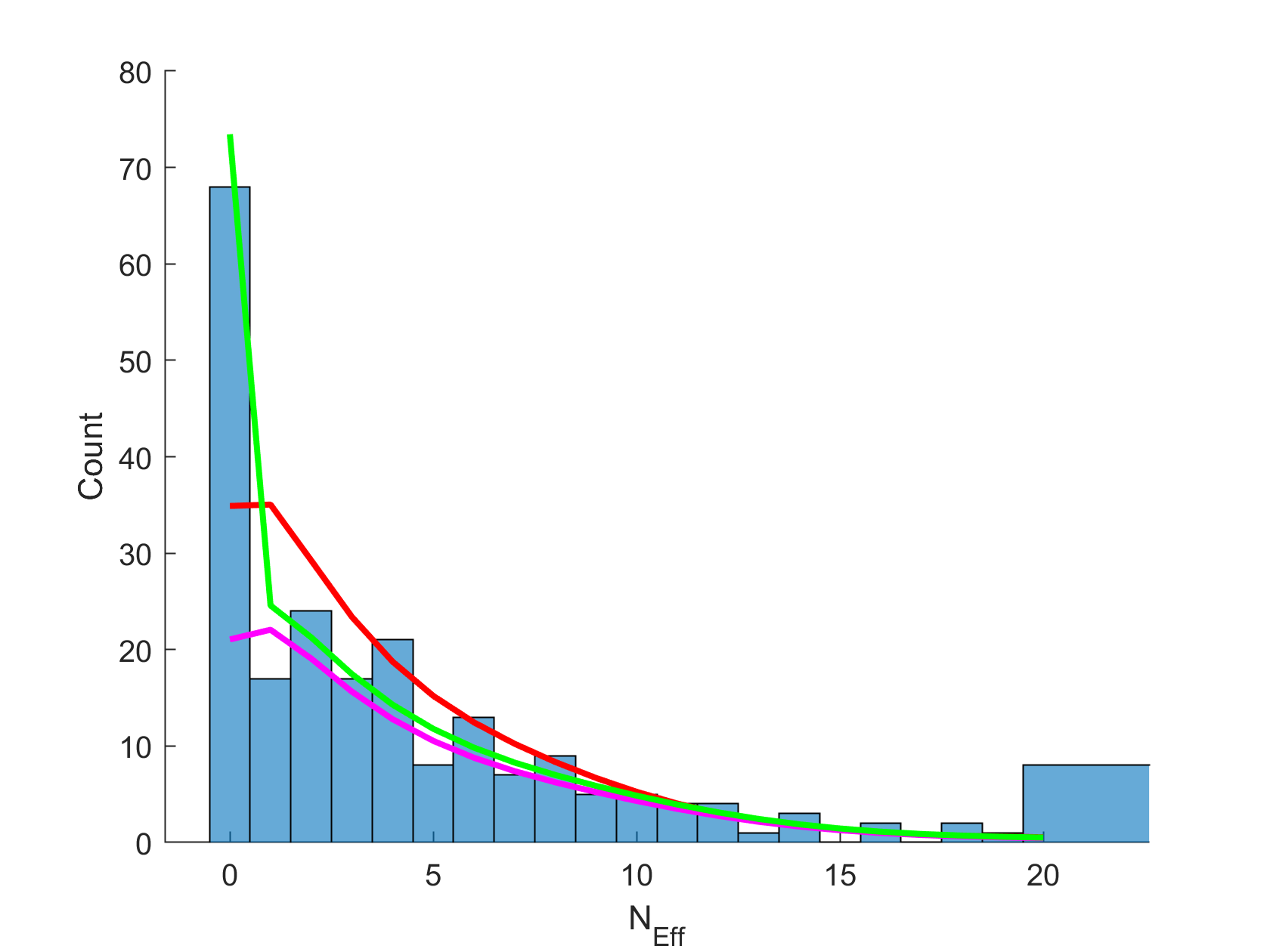}
\caption{\label{fig: P of N}
$N_\mathrm{Eff}$ histogram with three fits based on different models. The red curve represents a model with a constant NV density, $\rho$, and fits $N_\mathrm{Eff}$ data for all nanodiamonds. The magenta curve is a fit using the same model but only fitting data for fluorescent nanodiamonds. The green curve assumes a bimodal distribution of $\rho$ and utilizes all  $N_\mathrm{Eff}$ data. 
}

\end{figure}
}

\def \CsatCorrections{\begin{figure}[H]
\renewcommand\figurename{Supporting Information Figure}
\includegraphics[width = \textwidth ]{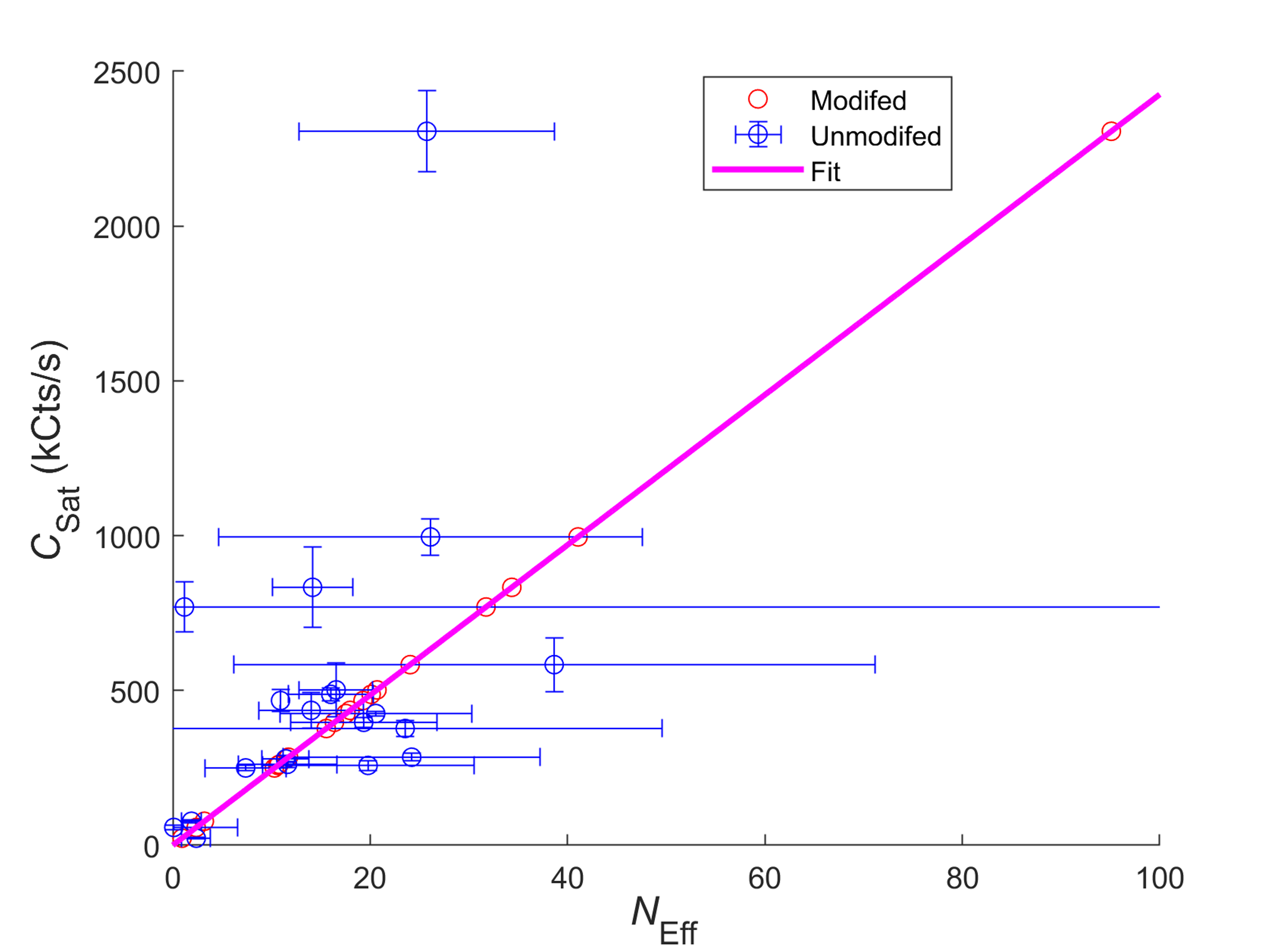}
\caption{\label{fig:Csat Corrections}
Plot of   $C_\mathrm{Sat}$ \textit{vs} $N_\mathrm{Eff}$ for the 20 nanodiamonds with unreliable $N_{Eff}$ values. Uncorrected data (blue), corrected data (red), $C_\mathrm{Sat}$ \textit{vs} $N_\mathrm{Eff}$ curve (magenta). 
}
\end{figure}
}

\def \HegihtVFrac{\begin{figure}[H]
\renewcommand\figurename{Supporting Information Figure}
\includegraphics[width = \textwidth ]{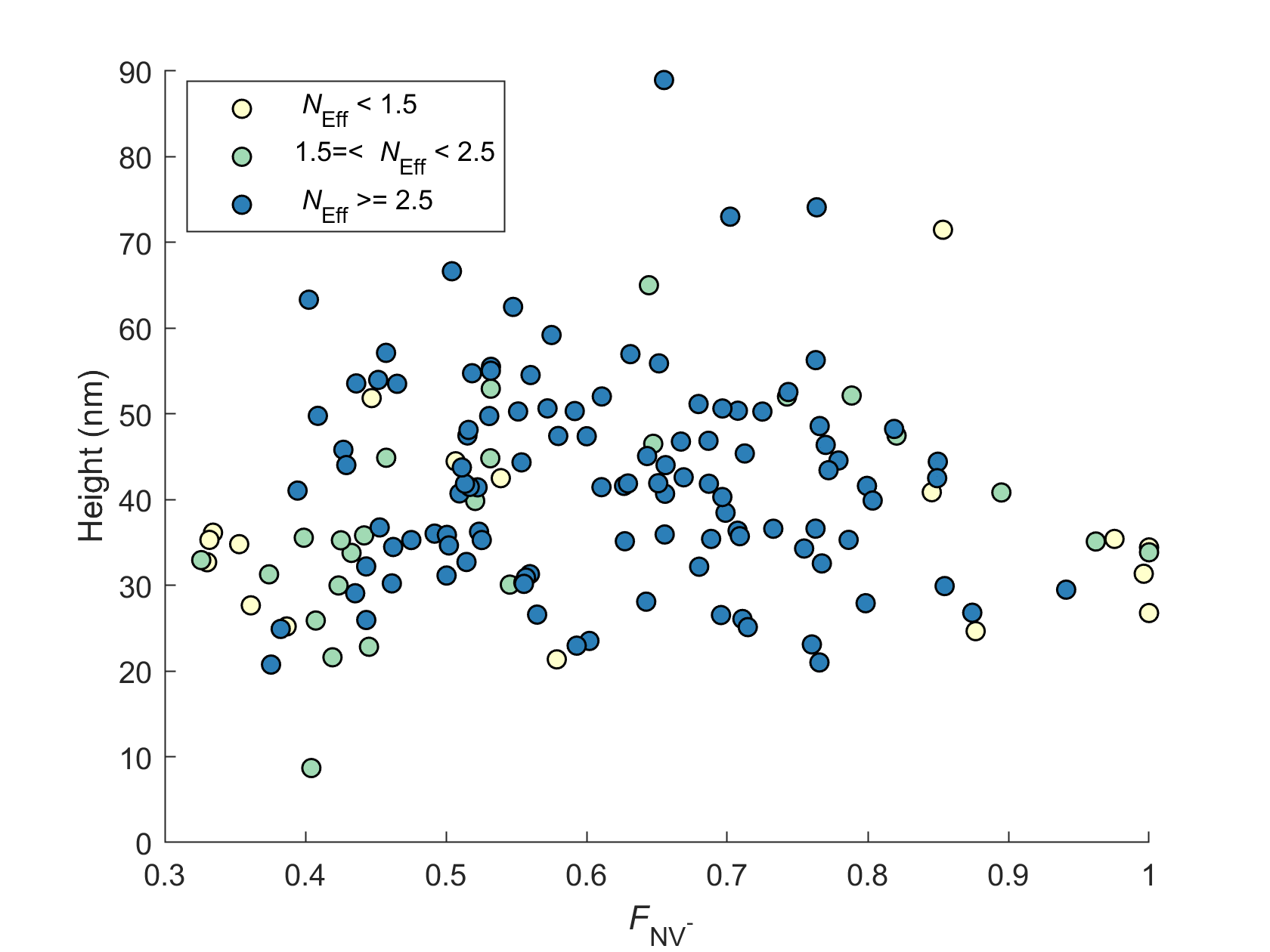}
\caption{\label{fig:Frav Vs Height}
Plot of   Height \textit{vs}$F_\mathrm{NV^{-}}$.
}
\end{figure}
}

\def \HeightVsT{\begin{figure}[H]
\renewcommand\figurename{Supporting Information Figure}
\includegraphics[width = \textwidth ]{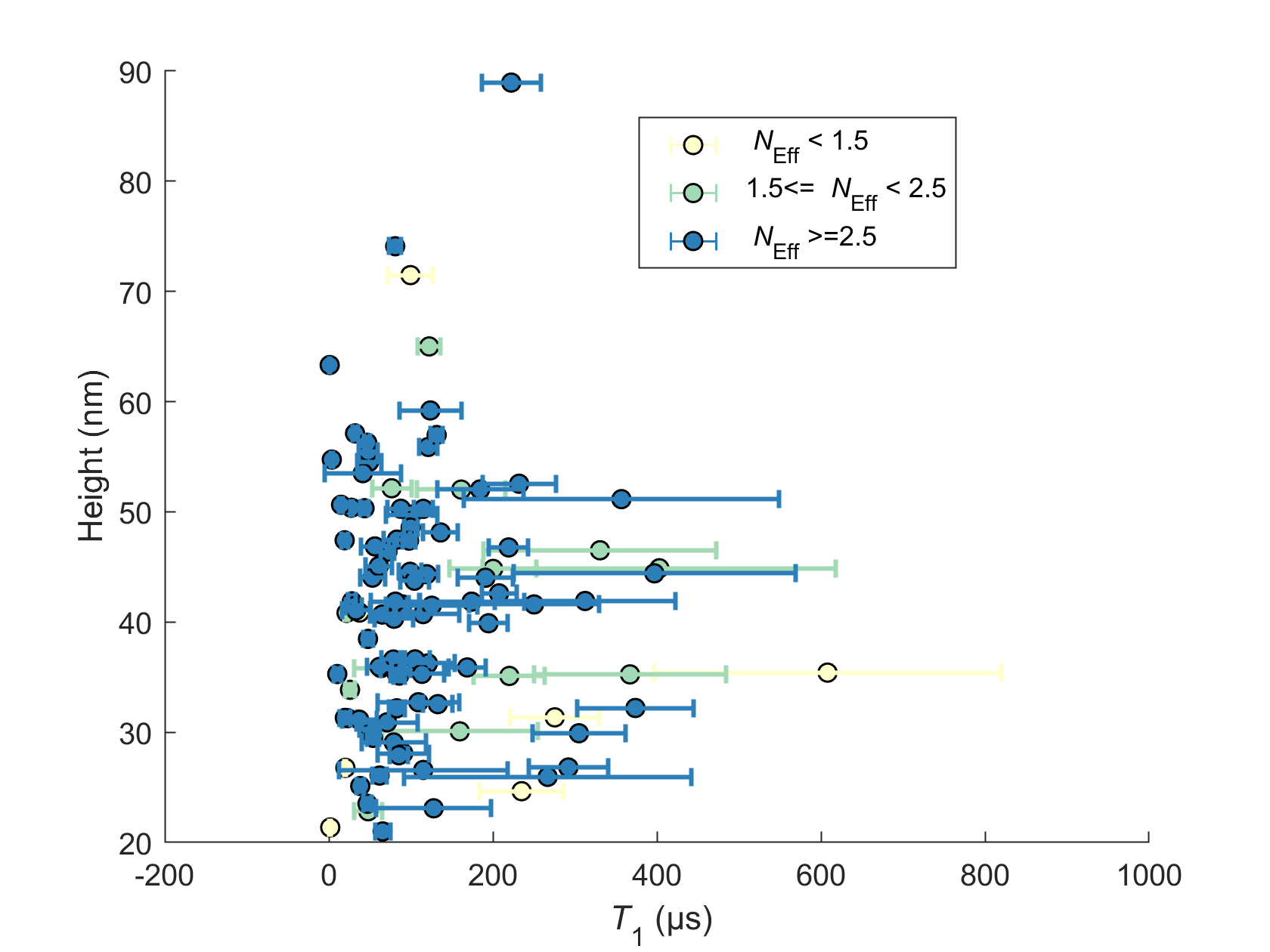}
\caption{\label{fig:T1 Vs Height}
Plot of Height  \textit{vs} $T_{1}$   for the measured nanodiamonds.
}
\end{figure}
}

\def \NVFracVsT{\begin{figure}[H]
\renewcommand\figurename{Supporting Information Figure}
\includegraphics[width = \textwidth ]{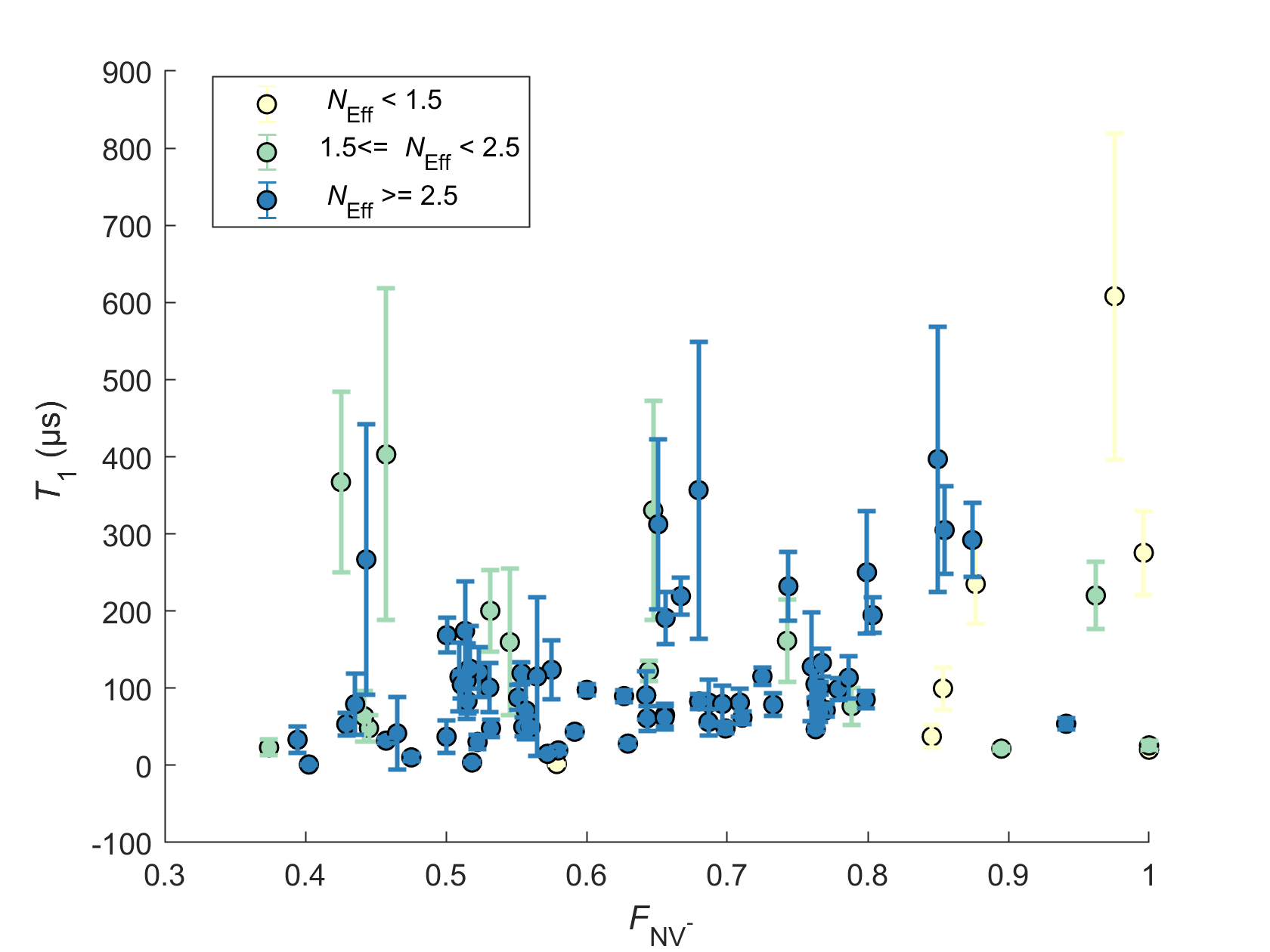}
\caption{\label{fig:Frav Vs T1}
Plot of $T_{1}$ \textit{vs} $F_\mathrm{NV^{-}}$ for the measured nanodiamonds.
}
\end{figure}
}

\def \PsatHistogram{\begin{figure}[H]
\renewcommand\figurename{Supporting Information Figure}
\includegraphics[width = \textwidth ]{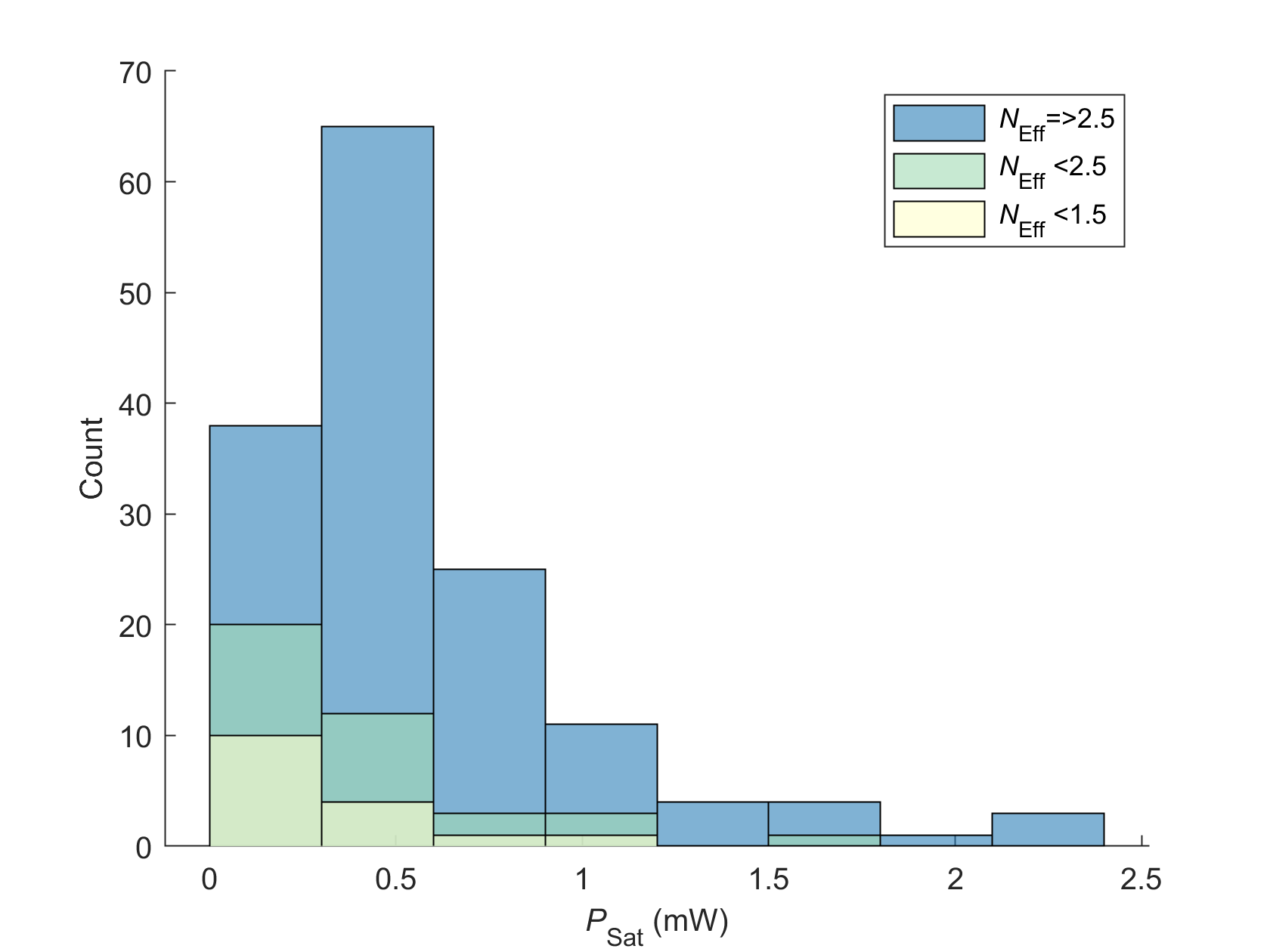}
\caption{\label{fig:PSat} Distributions of  $P_\mathrm{Sat}$ for the measured fluorescent nanodiamonds. 
}
\end{figure}
}

\maketitle{}


\section{Additional SEM, AFM, and PL Images} 
 \begin{figure}[H]
\renewcommand\figurename{Supporting Information Figure}
\includegraphics[scale=.5]{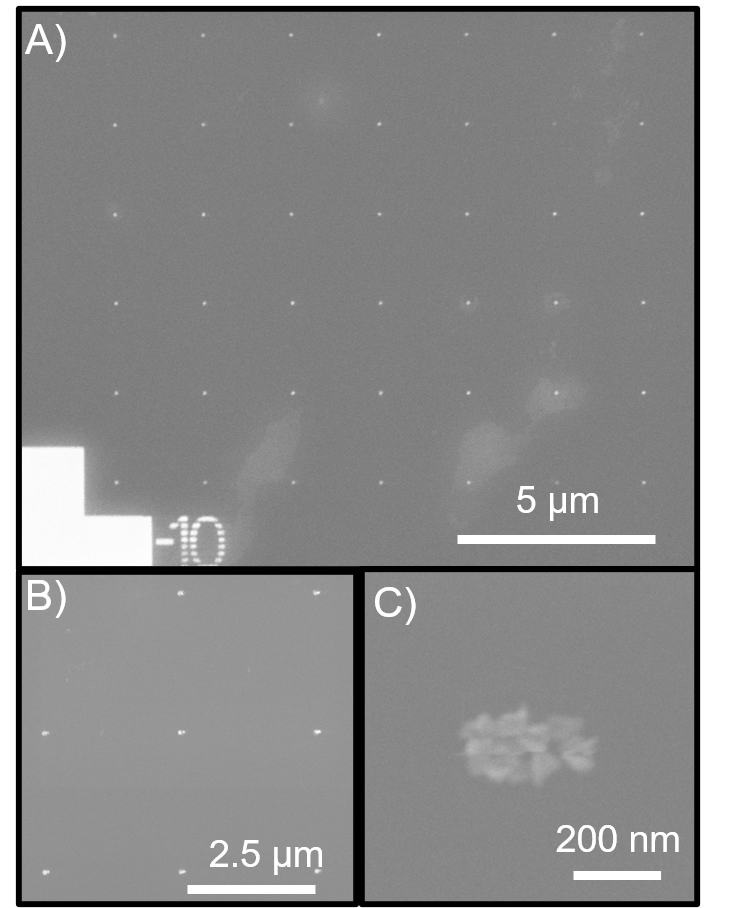}
\centering
\caption{\label{Large Template} Scanning electron microscope images of a 250 nm diameter template with assembles of nanodiamond clusters.  
}

\end{figure}

\begin{figure}[H]
\renewcommand\figurename{Supporting Information Figure}
\includegraphics[width = \textwidth ]{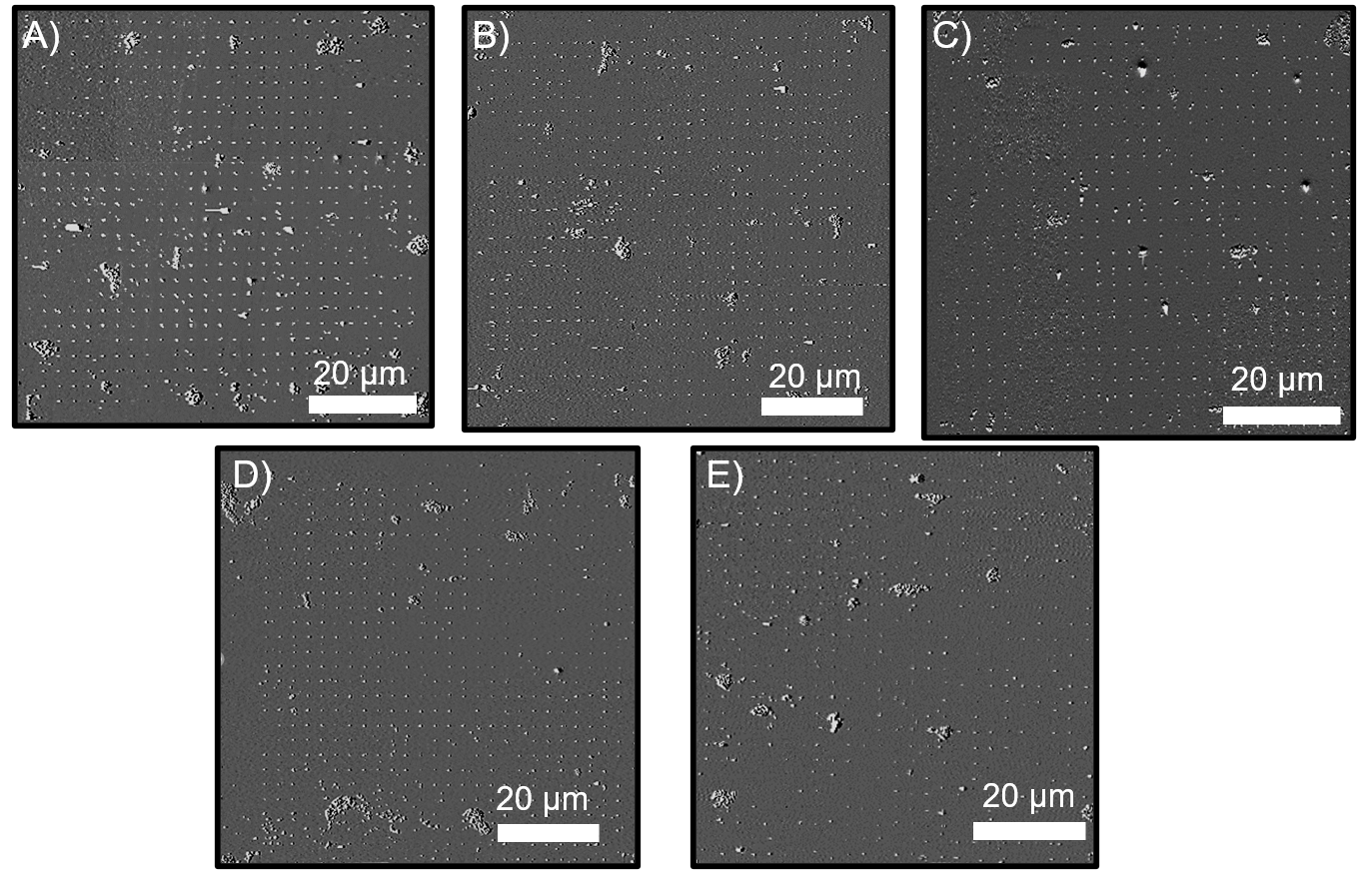}
\caption{\label{fig: AFM Areas}
AFM images mapping 26 $\times$ 26 nanodiamond arrays.  
(A) 48 nm diameter template, (B)  45 nm diameter template, (C)  42 nm diameter template, (D)  38  nm diameter template,(E)  35  nm diameter template.
}

\end{figure}

\begin{figure}{}
\renewcommand\figurename{Supporting Information Figure}
\includegraphics[width = \textwidth ]{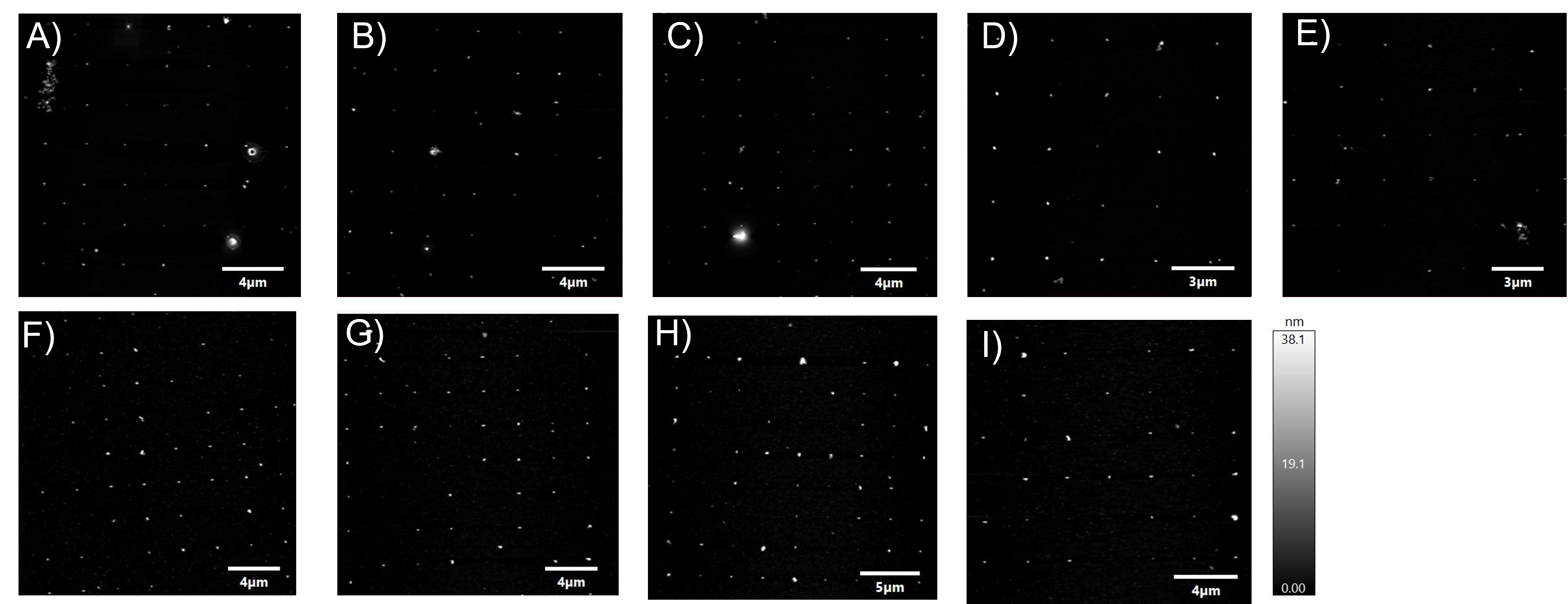}
\caption{\label{fig: AFM Areas Zoomed}
AFM images of the subarrays probed in automated measurements. 
}

\end{figure}

\begin{figure}[H]
\renewcommand\figurename{Supporting Information Figure}
\includegraphics[width = \textwidth ]{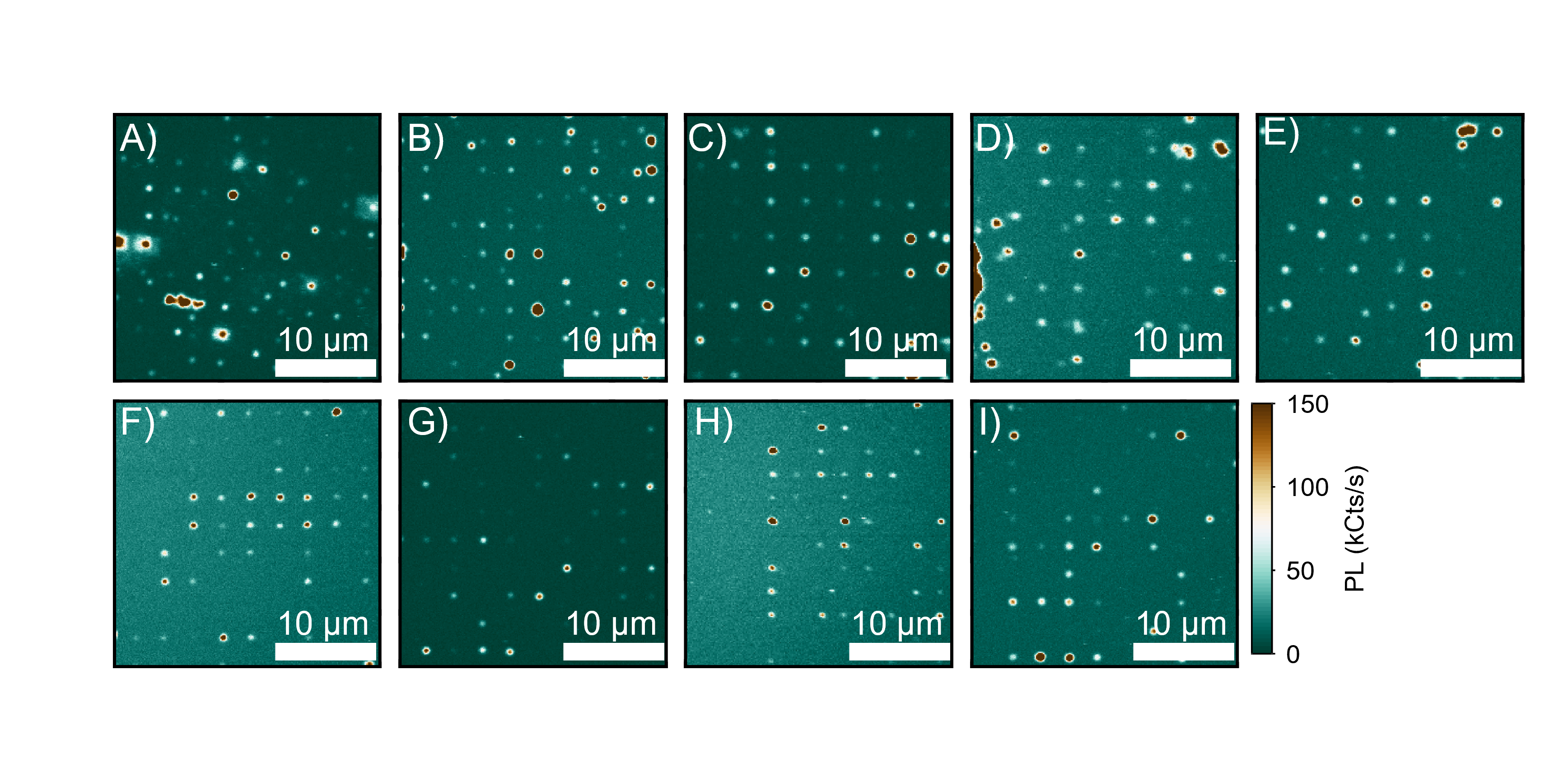}
\caption{\label{fig: PL Areas}
PL images of the subarrays probed in automated measurements. 
}

\end{figure}

\section{Results of Statistical Analysis }
\Corrlations{}
\CorrlationsPVal{}
\HegihtVFrac{}
\HeightVsT{}
\NVFracVsT{}

\section{Estimating NV Center Density}
\PofNFits{}

We estimate a rough NV center density in the final nanodiamonds using the specifications provided by Ada\'mas Nanotechnology. 
The nanodiamonds were irradiated with 2-3 MeV electrons to a dose of $10^{18}$ cm$^{-2}$  \cite{shenderova2017commercial}.
For this energy and dose, we assume a vacancy creation efficiency of order 1 cm$^{-1}$\cite{campbell2000radiation, hunt2000identification}.
While NV center creation efficiency depends on annealing temperature, nitrogen content,  and diamond size, recent studies employing similar synthesis conditions demonstrate conversions efficiencies on the order of 1-10\%\cite{mindarava2020efficient, capelli2019increased}.
Based on these assumptions, we estimate a final NV-center density of $\sim 10^{17}$ cm$^{-3}$, which is consistent with the value found in the vacancy limited region (Supporting Information Figure~S10).

From the best fit values of  $\rho_1$ and $\rho_2$ we can estimate the nitrogen distribution within the premilled diamond.
In the vacancy limited region, the lower bound of $\rho_1$ divided by the NV-center creation efficiency of 10\% (1\%) provides a lower limit on nitrogen density of 7.1 (71) ppm.  
In the nitrogen limited region, the upper bound of $\rho_1$ multiplied by the  NV-center creation efficiency provides an upper bound for nitrogen density of 0.25 (2.5) ppm.
The bounds for both regions are consistent with the reported distributions of accessible nitrogen within bulk high temperature high pressure synthetic diamonds\cite{shenderova2019synthesis, babich2009spatial, kanda1993inhomogeneous}.

\section{Relationship between Brightness and Emitter Number}
\CSatVN{}
\CsatCorrections{}
\PsatHistogram{}
By utilizing exclusively the 131 nanodiamonds where the ${N}_\mathrm{Eff}$ was deemed reliable, we fit our data using the linear function:
 $C_\mathrm{Sat} = A N_\mathrm{Eff} + B$ (Supporting Information Figure~S11).
 From this fit we find that individual  NV centers ($
N_\mathrm{Eff} = 1$) have $C_\mathrm{Sat}= A+B = 22.3 \pm 8.35$ kCts/s.
Each additional emitter adds $A= 20.72 \pm 1.13$ kCts/s.

\section{Nanodiamond Dispersion Spectra}
\Solution{}
The dispersion spectra are acquired by placing \SI{300}{\micro\liter}  of 1 mg/ml nanodiamond disperion loaded in a glass cuvette (SpectroCell R-403) in the confocal excitation path and illuminating with a 10x objective (0.3 NA 10X objective Nikon) at power of 4.7 mW before the objective.
We  perform a nonnegative matrix factorization analysis to compare this data with the spectral data of the assembled nanodiamonds.

\section{Core-Shell Model}
  \CoreShell{}
  To better understand the correlations between ${F}_\mathrm{NV^{-}}$,  ${T}_{1}$ and, size, we model the nanodiamonds as spherical particles with a core radius ($R_{C}$) and a shell thickness ($S$) such that $R_{C} + S$ is equal to the total radius of the sphere ($R$).
The shell volume ratio  ($V_{{S}/{R}}$)  is then given by:
  
   \begin{equation}\label{eq:S2C}
 V_{{S}/{R}}(R,S)= 1 - \frac{(R-S)^{3}}{(R)^{3}}
 \end{equation}

This relationship is plotted in Supporting Information Figure~S15 for a series of $S$ and $R$ values.
We can compare these volume ratios to the distribution of observed ${F}_\mathrm{NV^{-}}$ and ${T}_{1}$ values, assuming that NV centers residing in the shell region are more affected by the environment then defects in the core region, and hence have a lower ${F}_\mathrm{NV^{-}}$ and smaller ${T}_{1}$.
In this model, the likelihood of observing an NV center in the shell takes the form of a binomial distribution  with  a success probability $V_{{S}/{R}}$.
To perform this analysis, we sort the nanodiamonds by height and calculate the observed fraction of measurements that fall below a given threshold value for $F_\mathrm{NV^{-}}$ or $T_{1}$.
From this observed fraction and the number of measurements, we 
calculate the Wilson score interval as an estimate for the underlying binomial probability and corresponding uncertainty.
In Supporting Information Figure~S15, we plot the Wilson score intervals alongside the shell volume fraction for particular threshold values.
Although the uncertainties are large, the results are consistent with the interpretation that both $F_\mathrm{NV^{-}}$ and $T_{1}$ are reduced for NV centers located within a few nanometers of the nanodiamond surface.


 
\bibliography{Ref}